\documentclass[pra,twocolumn,superscriptaddress,showpacs,amsmath,amstex,amssymb,citeautoscript]{revtex4-1}

\usepackage[T1]{fontenc}
\usepackage[utf8]{inputenc}
\usepackage{lipsum}
\usepackage{amsmath}
\usepackage{amssymb}
\usepackage{bbm}
\usepackage{braket}
\usepackage{xcolor}
\usepackage{pifont}
\usepackage[mathscr]{euscript}
\usepackage[shortlabels]{enumitem}

\usepackage{graphicx}
\usepackage{lipsum}
\allowdisplaybreaks
\usepackage{float}
\usepackage{graphicx}
\usepackage{dsfont}
\usepackage{comment}
\usepackage[colorlinks=true]{hyperref}  
\hypersetup{
    bookmarks=true,         
    unicode=false,          
    pdftoolbar=true,        
    pdfmenubar=true,        
    pdffitwindow=false,     
    pdfstartview={FitH},    
    pdftitle={Probing superconductivity with tunneling spectroscopy in rhombohedral graphene},    
    pdfauthor={Sedov and Scheurer},     
    pdfsubject={},   
    pdfcreator={},   
    pdfproducer={}, 
    pdfkeywords={} {} {}, 
    pdfnewwindow=true,      
    colorlinks=true,       
    linkcolor=blue, 
    citecolor=blue,        
    filecolor=magenta,      
    urlcolor=blue           
}

\newcommand{\equref}[1]{Eq.~(\ref{#1})}

\newcommand{\figref}[1]{Fig.~\ref{#1}}
\newcommand{\refcite}[1]{Ref.~\onlinecite{#1}}

\newcommand{\tableref}[1]{Table~\ref{#1}}
\newcommand{\appref}[1]{Appendix~\ref{#1}}

\newcommand{\nf}{n_{\mathrm{F}}}
\newcommand{\pdagger}{{\phantom{\dagger}}}
\newcommand{\diff}{\mathrm{d}}

\renewcommand{\approx}{\simeq}

\renewcommand{\Im}{\text{Im}}

\renewcommand{\vec}[1]{\mathbf{#1}}

\definecolor{wrongultramarine}{rgb}{1,0.5,0}

\begin{document}

\title{Probing superconductivity with tunneling spectroscopy in rhombohedral graphene}

\author{Denis Sedov}
\affiliation{Institute for Theoretical Physics III, University of Stuttgart, 70550 Stuttgart, Germany}

\author{Mathias S.~Scheurer}
\affiliation{Institute for Theoretical Physics III, University of Stuttgart, 70550 Stuttgart, Germany}

\begin{abstract}
Motivated by experiments on rhombohedral tetralayer graphene showing signs of superconductivity emerging from a valley-polarized normal state, we here analyze theoretically how scanning tunneling spectroscopy can be used to probe the superconducting order parameter of the system. To describe different pairing scenarios on equal footing, we develop a microscopic tunneling approach that can capture arbitrary, including finite-momentum, superconducting order parameters and low-symmetry normal-state Hamiltonians. Our analysis shows that the broken time-reversal symmetry in a single valley leads to unique features in the weak-tunneling regime that are different for commensurate and incommensurate Cooper pair momenta. We further uncover an unconventional spatial dependence of the Andreev conductance, allowing to distinguish between three topologically distinct classes of single-$\vec{q}$ pairing states in the system, and compute the signatures of a competing translational-symmetry breaking three-$\vec{q}$ ``moiré superconductor''. 
\end{abstract}

\maketitle

\begin{figure}[t]
    \includegraphics[width=0.48\textwidth]{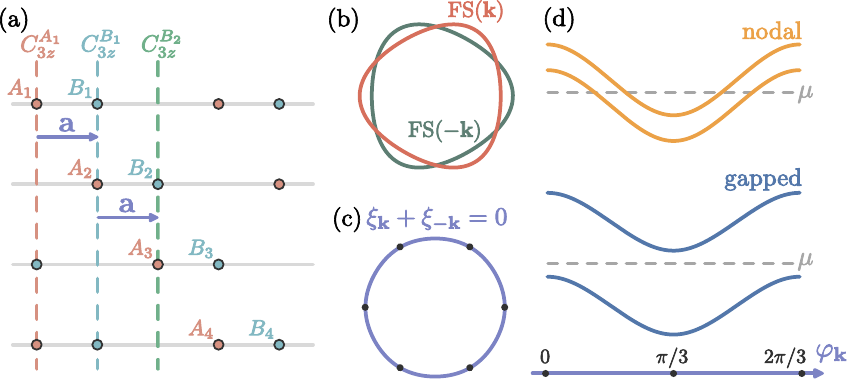}
    \caption{a) Schematic illustration of the geometry of RTG; $A_n, B_n$ are sublattice atoms of the $n$th graphene layer. b) Fermi surfaces of electrons and holes. c) Surface in $\vec{k}$-space defined by the expression $\xi_{\mathbf{k}} + \xi_{-\mathbf{k}} = 0$; points on the surface correspond to the saddle and maximum/minimum points of the Bogoliubov excitations spectrum. d) Bogoliubov excitation spectrum along the contour presented in c) in the nodal and gapped pairing regimes.}
    \label{fig:geometry_and_crossover}
\end{figure}

In recent years, multilayer graphene systems have been established as a fascinating route to complex correlated physics \cite{review1,review2,review3,TwistedNonTwisted,HalfQuarterMetals,SCTrilayer}. In addition to other interaction-induced phases, such as spontaneous polarization of the different internal ``flavor'' degrees of freedom \cite{HalfQuarterMetals,SCTrilayer,Wong_2020,Zondiner_2020,Microwave}, the fact that these systems host superconducting states \cite{SCTrilayer,SuperconductivityTBG} has been a particularly strong driving force of research. Although tunneling experiments hint at pairing beyond the simple BCS paradigm in twisted multilayer graphene \cite{Oh_2021,TunnelingPerge,PhysRevB.106.104506,PhysRevLett.130.216002,PhysRevLett.131.016003,PhysRevB.107.L020502,vestigialSC,BandOffDiagonalPairing,PhysRevB.111.064515}, the form of the superconducting order parameter and the pairing glue are still subject of debate. 
After the discovery \cite{SCTrilayer} and subsequent theoretical study \cite{PhysRevLett.127.187001,PhysRevLett.127.247001,PhysRevB.105.134524,ShubhayuPairing,PhysRevB.107.104502,PhysRevB.107.L161106,PhysRevLett.130.146001,PhysRevB.108.045404,PhysRevB.105.L081407,PhysRevB.108.134503,2024arXiv240617036D} of superconductivity in rhombohedral stacks of graphene, see \figref{fig:geometry_and_crossover}a, very recently, an experiment \cite{valleySCExp} revealed a rather remarkable behavior in rhombohedral tetralayer graphene (RTG); here, superconductivity seems to be born out of a valley-polarized normal state, i.e., the normal-state Fermi surfaces are exclusively located in the vicinity of the $\vec{K}$ and not the $\vec{K}'$ point (or vice versa). For one of the two valley-polarized superconductors, the spin degree of freedom is also likely polarized. Multiple works studying the energetics and properties of superconductivity in valley-imbalanced systems \cite{scammell_theory_2022,chou2024intravalleyspinpolarizedsuperconductivityrhombohedral,geier2024chiraltopologicalsuperconductivityisospin,yang2024topologicalincommensuratefuldeferrelllarkinovchinnikovsuperconductor,qin2024chiralfinitemomentumsuperconductivitytetralayer,2025arXiv250219474P,2025arXiv250217555Y,TheEnergeticsPaper,2024arXiv241109664J,2025arXiv250219474P,2025arXiv250305697M,Kim_2025,PhysRevLett.132.046003,PhysRevResearch.6.043240} demonstrate that the lack of orbital time-reversal symmetry (TRS) in the normal state can lead to exotic physical properties.

As a clear experimental identification of the form of superconductivity is typically very challenging,
this work explores theoretically how scanning tunneling spectroscopy (STS) can be used to gain crucial information about pairing in valley-polarized RTG. This is a complex endeavor because a sufficiently large valley polarization is not only expected to lead to $\vec{q}=0$ intravalley pairing, i.e., Cooper pairs with a commensurate center-of-mass momentum $2\vec{K}$, but also to states with incommensurate momentum $2\vec{K}+\vec{q}$, where $\vec{q} \neq 0$ \cite{scammell_theory_2022}. Furthermore,  the off-resonant band energies of the paired electrons lead to nodal or fully gapped regimes \cite{chou2024intravalleyspinpolarizedsuperconductivityrhombohedral,geier2024chiraltopologicalsuperconductivityisospin,yang2024topologicalincommensuratefuldeferrelllarkinovchinnikovsuperconductor,qin2024chiralfinitemomentumsuperconductivitytetralayer,2025arXiv250219474P,2025arXiv250217555Y,2025arXiv250219474P,TheEnergeticsPaper}, depending on the size of the order parameter, and, for $\vec{q}=0$, one has to distinguish pairing in three distinct irreducible representations (IRs) $A$, $E$, $E^*$ of the three-fold rotational symmetry $C_{3z}$; at $\vec{q}\neq 0$ these states can still be distinguished based on their Chern number. Finally, \refcite{TheEnergeticsPaper} showed that apart from the aforementioned 1-$\vec{q}$ states, also 3-$\vec{q}$ states are possible which restore $C_{3z}$ but break translational symmetry.  We here demonstrate that---taking advantage of the unique features of intravalley pairing---a combination of weak- and strong-tunneling measurements at different high-symmetry positions will provide key insights about which of the above-mentioned pairing states is present in the system. 


\vspace{1em}
\textit{General formalism}---To model STS experiments in RTG systematically, we incorporate and generalize the Hamiltonian approach to tunneling based on the Keldysh formalism~\cite{hamiltonian_approach_to_NS} to the case with arbitrary momentum dependent superconducting (SC) order parameters $\Delta_{\vec{k},-\vec{k}'}$, coupling electronic states at in general non-zero net momentum, $\vec{k} - \vec{k}' \neq 0$, and without TRS in the multi-component normal-state Bloch Hamiltonian $h_{\vec{k}}$. More explicitly, the target system, here RTG, is described by $\hat{H}_{c} = \sum_{\mathbf{k}} (c^\dagger_{\mathbf{k},\uparrow} h_{\mathbf{k}} c_{\mathbf{k},\uparrow} - c_{-\mathbf{k},\downarrow} \cdot h_{-\mathbf{k}}^{\mathrm{T}} \cdot c^\dagger_{-\mathbf{k},\downarrow}) + \sum_{\mathbf{k},\mathbf{k}'} ( c^\dagger_{\mathbf{k},\uparrow} \cdot \Delta_{\mathbf{k},-\mathbf{k}'} \cdot c^{\dagger}_{-\mathbf{k}',\downarrow} + \mathrm{H.c.})$, where $h_{\mathbf{k}}$ is the continuum 8-band model~\cite{PhysRevB.82.035409, Model_Koshino} of RTG and $c^\dagger_{\mathbf{k},\sigma}$ is a creation operator of electrons in the $\mathbf{K}$-valley with momentum $\mathbf{k}$ and spin $\sigma$. We assume that the tunneling is local with respect to the lead degrees of freedom, $\hat{H}_t =  \sum_{\mathbf{k},\sigma} f^{\dagger}_{\sigma}(0) t^{\dagger}_{\mathbf{k}} c_{\mathbf{k},\sigma} + \mathrm{H.c.}$, where $f_{\sigma}^\dagger(0)$ creates an electron at the tip of the lead and $t_{\vec{k}}$ are the tunneling matrix elements. We also assume that both systems are kept at the same temperature and the difference between their chemical potentials is controlled by the STS bias voltage $eV$. Thus, we neglect any effects of the distribution function renormalization.

The tunneling current can be divided into four distinct terms with different physical interpretations (\cite{hamiltonian_approach_to_NS}, SI). While we always include all of them in our numerics, for our discussion here, only two of them are crucial. First, in the weak-tunneling regime, when tunneling amplitudes are much smaller than characteristic energies of the systems (e.g.~bandwidth, hopping constants), the current is mostly determined by the famous expression, $I_{\text{weak}} \sim \int d\omega \rho_{f,\text{pp}}(\omega - eV) \sum_{\mathbf{k},\mathbf{k}'} t^\dagger_{\mathbf{k}} \rho_{c,\text{pp}}(\omega, \mathbf{k}, \mathbf{k}') t_{\mathbf{k}'} \left[ \nf(\omega - eV) - \nf(\omega) \right] + o(\| t \|^2)$, where $\rho_{f/c,pp}$ are bare particle spectral functions. The corresponding tunneling conductance, $G(eV) = \partial I / \partial (eV)$, is then proportional to the local density of states (LDOS). Therefore, weak-tunneling allows to probe spectral features of the Bogoliubov excitations. Second, when tunneling amplitudes become large enough, the Andreev contribution to the current become non-negligible. It is proportional to the SC order parameter $\Delta_{\mathbf{k},-\mathbf{k}'}$ and reads as
\begin{align}\label{AndreevCurrent}
    &I_{\mathrm{A}} = \frac{4\pi e}{\hbar} \int d\omega \biggl| \sum_{\mathbf{k},\mathbf{k}'} t^{\dagger}_{\mathbf{k}} G_{cc,\mathrm{ph}}^r(\omega,\mathbf{k},\mathbf{k}') t^*_{-\mathbf{k}'} \biggr|^2  \\
    &\times \rho_{f}^{\mathrm{pp}}(\omega - eV) \rho_{f}^{\mathrm{hh}}(\omega + eV) \left[ \nf(\omega - eV) - \nf(\omega + eV) \right], \nonumber
\end{align}
where $G_{cc,\mathrm{ph}}^r \propto \Delta_{\mathbf{k},-\mathbf{k}'}$ is the particle-hole (anomalous) part of the retarded Greens function of the RTG in the Nambu basis. The Andreev current is generated in the process in which electrons tunnel from the lead to the SC, then by the Andreev scattering are converted to a hole tunneling back to the lead. This is the only contribution to the current that does not generally vanish when $|eV|$ lies within the gap of the Bogoliubov excitations. This feature along with its proportionality to the SC order parameter makes the strong tunneling regime a useful tool to extract information about the symmetry of the SC from STS experiments.

\vspace{1em}
\textit{Weak tunneling}---We begin our analysis in the weak-tunneling regime, where the current is dominated by $I_{\text{weak}}$ following the LDOS. We further first focus on $\vec{q}=0$  intravalley pairing (commensurate momentum $2\vec{K}$) in valley-polarized RTG and only keep the band crossing the Fermi level. Then superconductivity is described by $H_{1\text{b}} = \sum_{\mathbf{k},\sigma} \xi_{\mathbf{k}}d^\dagger_{\mathbf{k},\sigma} d^\pdagger_{\mathbf{k},\sigma} + H_{\Delta}$, where $H_{\Delta}=\sum_{\mathbf{k}} ( d^{\dagger}_{\mathbf{k},\uparrow} \Delta_{\mathbf{k}} d^{\dagger}_{-\mathbf{k},\downarrow} + \mathrm{H.c.})$ and $d^\dagger_{\mathbf{k},\uparrow\downarrow}$ are electronic creation operators in the active band hosting superconductivity.

To gain intuition, we first consider the corresponding Bogoliubov spectrum, given by $E_{\mathbf{k}}^{\pm} = 1/2 (\xi_{\mathbf{k}} - \xi_{-\mathbf{k}} \pm \sqrt{(\xi_{\mathbf{k}} + \xi_{-\mathbf{k}})^2 + 4 |\Delta_{\mathbf{k}}|^2})$. 
Broken TRS, $\xi_{\mathbf{k}} \neq \xi_{-\mathbf{k}}$, see \figref{fig:geometry_and_crossover}b, results in two pairing regimes---nodal and fully gapped---controlled by the magnitude $|\Delta_{\mathbf{k}}|$ of the order parameter, which we set to be $\vec{k}$ independent, $\Delta_0 = |\Delta_{\mathbf{k}}|$, for simplicity of the following discussion. Key features can be understood by considering the behavior of the excitations on the contour $\xi_{\mathbf{k}} + \xi_{\mathbf{-k}} = 0$ in $\mathbf{k}$-space presented in the \figref{fig:geometry_and_crossover}c. Due to $C_{3z}$-symmetry, implying $\xi_{C_{3z}\mathbf{k}} = \xi_{\mathbf{k}}$, the three saddle and minimum (maximum) points of $E_{\mathbf{k}}^{+(-)}$ are located on this contour, and when $\Delta_0$ becomes large enough, the minimum of $E^{+}_{\mathbf{k}}$ crosses the Fermi level, see  \figref{fig:geometry_and_crossover}d. 

\begin{figure}[b]
    \centering
    \includegraphics[width=0.48\textwidth]{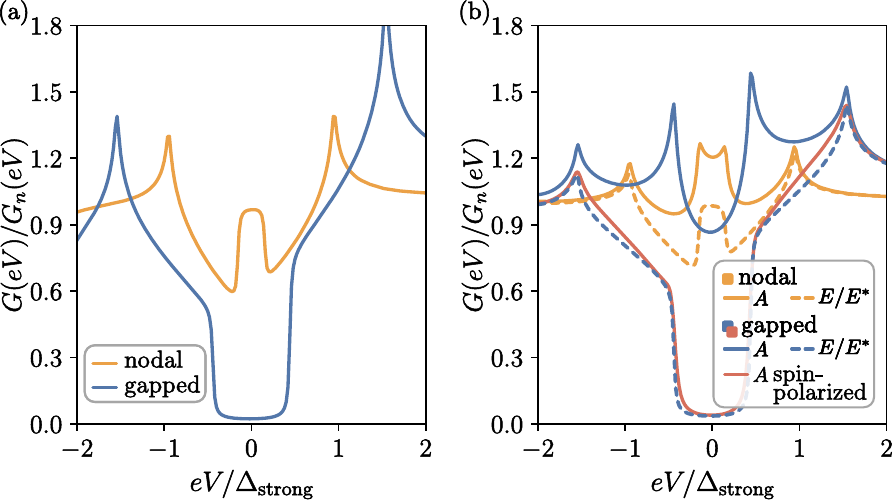}
    \caption{Normalized tunneling conductance for tunneling into the $A_1$ sublattice in a) weak and b) strong tunneling regimes. Weak tunneling conductance is the same for all IRs of the order parameter. Life-time broadening $\eta / \Delta_{\mathrm{nodal}} = 0.0375$, temperature $T / \Delta_{\mathrm{nodal}} = 0.0375$, $|t^2_{\mathrm{weak}} \rho_{f,n}(0) \rho_{c,n}(0)| = 0.25\times 10^{3}, |t^2_{\mathrm{strong}} \rho_{f,n}(0) \rho_{c,n}(0)| = 2.5$, where $\rho_{f/c,n}(0)$ are the normal LDOS of the systems.}
    \label{fig:weak_strong_pairing}
\end{figure}

The discussed behavior is visible in the weak-tunneling conductance, shown in \figref{fig:weak_strong_pairing}a. Sharp peaks correspond to divergencies of the DOS related to the aforementioned three saddle points of $E_{\mathbf{k}}^{\pm}$. 
In contrast to systems with preserved TRS, these peaks do not appear at the edges of the hard gap, but above it. In the gapped pairing case, in the vicinity of the hard gap energy, which is smaller than the order-parameter magnitude $\Delta_0$ \cite{BandOffDiagonalPairing}, the spectrum $E_{\mathbf{k}}^{\pm}$ behaves quadratically in $\vec{k}$. This leads to a steplike behavior of the tunneling conductance instead of the usual ``coherence peaks''. The nodal regime is characterized by a finite-height plateau near the Fermi energy, almost coinciding with the normal tunneling conductance.  We note that additional spin polarization in the normal state does not change these conclusions.

\vspace{1em}
\textit{Strong tunneling}---The weak-tunneling regime gives access to unique features associated with broken TRS in RTG, but it is not sensitive to the phase of $\Delta_{\vec{k}}$ and, thus, the IR of the order parameter. This is different in the strong-tunneling regime, where $I_{\text{A}}$ in \equref{AndreevCurrent} can become dominant. To show this, we choose a gauge in which $d_{\mathbf{k}}$ transform trivially under three-fold rotations around $A_1$ (see \figref{fig:geometry_and_crossover}a), without acquiring a phase, $C_{3z}^{A_1}: d_{\mathbf{k}}^\dagger \to d_{C_{3z}\mathbf{k}}^\dagger$. We then associate the phase factor $w = 1, e^{2\pi i / 3}, e^{-2\pi i /3}$ in the action of $C_{3z}^{A_1}$ on the order parameter, $\Delta_{\mathbf{k}} \to \Delta_{C_{3z}^{-1}\mathbf{k}} = w \Delta_{\mathbf{k}}$, with IRs $A, E, E^{*}$, respectively \footnote{We here neglect an additional reflection symmetry $\sigma_v$ for simplicity of the discussion.}. Although other choices are also possible (cf.~other rotation axes below), we here use this convention for the following reason: In general, one can split the Berry curvature of the superconductor into three contributions, $\Omega=\Omega_{\text{SC}} + \Omega_{\text{b}} + \delta \Omega$ (see SI). Here, $\Omega_{\text{SC}}$ is the Berry curvature of the Bogoliubov Hamiltonian of the single-band model $H_{1\text{b}} = \sum_\vec{k} \Psi_{\vec{k}}^\dagger h_{\text{BdG}}(\vec{k}) \Psi_{\vec{k}}^\pdagger$, with Nambu spinor $\Psi_{\vec{k}}$, i.e., $\Omega_{\text{SC}} = \vec{g}_{\vec{k}} \cdot (\partial_x \vec{g}_{\vec{k}} \times \partial_y \vec{g}_{\vec{k}})/(2 |\vec{g}_{\vec{k}}|^3)$ where $h_{\text{BdG}}(\vec{k})=E_0(\vec{k}) \tau_0 + \vec{g}_{\vec{k}} \cdot (\tau_x,\tau_y,\tau_z)^T$. Furthermore, $\Omega_{\text{b}}$ is the contribution from the Berry curvature of the bands we study pairing in and $\delta \Omega$ a term involving both, which restores gauge invariance; while $\Omega_{\text{b}}$ and $\Omega$ are gauge-invariant separately, $\Omega_{\text{SC}}$ is not, since $d_{\mathbf{k}} \rightarrow e^{i\varphi_{\mathbf{k}}} d_{\mathbf{k}}$ will add a $\mathbf{k}$-dependent phase to $\Delta_{\mathbf{k}}$. It is straightforward to show that, for the Bloch bands of a minimal two-band model \cite{Model_Koshino} of RTG, the contributions of $\Omega_{\text{b}}$ and $\delta \Omega$ to the Chern number $C$ of the superconductor cancel out with the contribution coming from higher energies beyond the continuum model in the above-mentioned gauge where $C_{3z}^{A_1}$ acts trivially (see SI). We then simply have $C = \frac{1}{2\pi}\int \diff^2 \vec{k} \, \Omega_{\text{SC}}$ such that the $A$ state with $\Delta_{\mathbf{k}} = \Delta_0$ and $E$/$E^*$ state with $\Delta_{\mathbf{k}} = \Delta_0 e^{\pm i\varphi_{\mathbf{k}}}$ are topologically trivial, $C=0$, and non-trivial with $C=\pm 1$, respectively.

For these order parameters and taking a spin-unpolarized normal state, we show the conductance in the strong-tunneling regime in \figref{fig:weak_strong_pairing}b when tunneling exclusively into the $A_1$ sublattice. We can see that the result for $E/E^*$ now differs significantly from that of $A$, since the former only exhibits slight modifications in the normalized conductance compared to weak tunneling, while the latter behaves rather differently. Most notably, in the gapped regime, only $A$ exhibits a sizable Andreev signal below the gap. This can be straightforwardly understood by noting that the $\vec{k}$ sum in \equref{AndreevCurrent} vanishes by symmetry. This results from the fact that tunneling and thus $t_{\vec{k}}$ respect $C_{3z}^{A_1}$ while $\Delta_{\vec{k}}$ transforms non-trivially under it for $E$ and $E^*$ pairing. 

Apart from $C_{3z}^{A_1}$, RTG has two additional rotational symmetries, $C_{3z}^{B_1}$ and $C_{3z}^{B_2}$, as shown in \figref{fig:geometry_and_crossover}a. As they are related by lattice translational symmetry, it holds $C_{3z}^{M}: d_{\mathbf{k}}^\dagger \to \gamma_M e^{i \mathbf{a}_{M}(\mathbf{k} - C_{3z}\mathbf{k})} d_{C_{3z}\mathbf{k}}^\dagger$ with $\mathbf{a}_{M} = 0, \vec{a}, 2\vec{a}$ and phases $\gamma_M = 1, e^{-2\pi i/3}, e^{2\pi i /3}$ for $M = A_1, B_1, B_2$; the phases cancel out in an intervalley superconducting order parameter, which thus transforms identically under all of them. In contrast, this is not the case for our intravalley superconductor which carries finite, lattice-commensurate momentum, $C_{3z}^{M}: \Delta_{\mathbf{k}} \to \gamma_M^{2} \Delta_{C_{3z}^{-1}\mathbf{k}} = \gamma_M^2 w \Delta_{\mathbf{k}}$. Therefore, e.g., the $A$ state defined above transforms under IR $E$ and $E^*$ with respect to $C_{3z}^{B_1}$ and $C_{3z}^{B_2}$. This is of direct experimental relevance since changing the tip position from $A_1$ to $B_{1,2}$ will break $C_{3z}^{A_1}$ symmetry in $t_{\vec{k}}$ and, instead, $C_{3z}^{B_{1,2}}$ will be preserved. As such, moving the tip position will lead to a cyclic permutation of which of the three superconductors with Chern number $C=0,\pm 1$ exhibits the Andreev peak. Consequently, Andreev spectroscopy at multiple tip positions will yield crucial, phase-sensitive information about the superconductor, which is unique to intravalley pairing. 
Note that the magnitude of the normal tunneling conductance also changes since the lowest band is mostly polarized in the $A_1$-$B_4$ subspace. 

We finally mention that additional full spin polarization of the normal state will generally lead to a suppression of the Andreev peak (in the local tunneling approximation), which we also show in \figref{fig:weak_strong_pairing}b using the $A$ state as an example: as follows again from the $\vec{k}$ sum in \equref{AndreevCurrent}, the Andreev signal has to vanish at $eV=0$ due to the fermionic anti-symmetry, $\Delta_{\mathbf{k}} = -\Delta_{-\mathbf{k}}$, but is finite for $eV\neq 0$. This is to be contrasted with spin polarization in intervalley single-band pairing, where TRS implies $I_{\text{A}} = 0$ for \textit{all} energies (see SI).


\begin{figure}[t]
    \centering
    \includegraphics[width=0.48\textwidth]{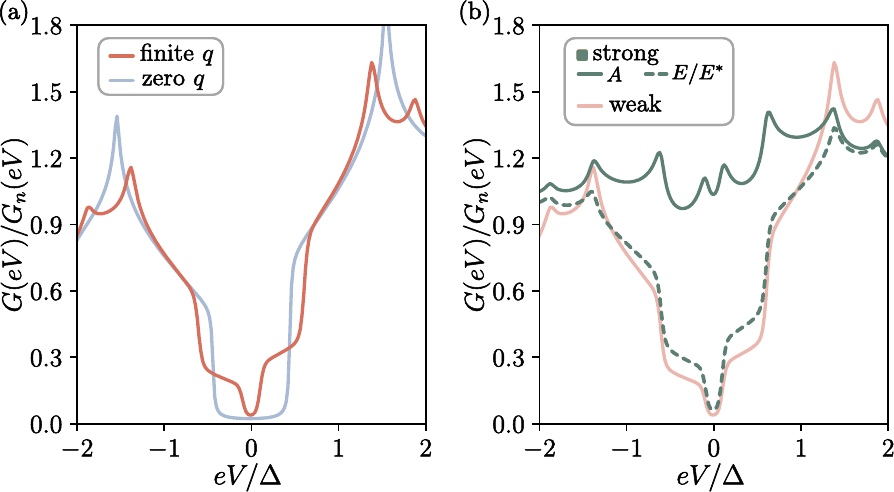}
    \caption{Tunneling conductance of the finite-momentum 1-$\vec{q}$ state. a) Comparison of the weak tunneling conductance between zero-$q$ and finite-$q$ SC. b) Comparison between weak and strong tunneling conductance of the finite-$q$ SC for different IRs of the SC order parameter. Weak tunneling conductance is independent of the IRs of the order parameter. $\eta / \Delta_{\mathrm{nodal}} = 0.0375$, $T / \Delta_{\mathrm{nodal}} = 0.0375$, $|t^2_{\mathrm{weak}} \rho_{f,n}(0) \rho_{c,n}(0)| = 0.25\times 10^{3}, |t^2_{\mathrm{strong}} \rho_{f,n}(0) \rho_{c,n}(0)| = 2.5$, $\mathbf{q}/\braket{k_{\mathrm{F}}} = 0.1 (1,0)$, where $\braket{k_{\mathrm{F}}}$ is the angle-averaged absolute value of the Fermi vector.}
    \label{fig:finite_q}
\end{figure}

\vspace{1em}
\textit{Finite momentum pairing}---Although $C_{3z}^M$ can pin $\vec{q}$ to zero for an extended set of parameters \cite{scammell_theory_2022}, broken TRS can also lead to pairing at incommensurate momenta, i.e., $\vec{q} \neq 0$ in $H_{\Delta} = \sum_{\mathbf{k}} d^\dagger_{\mathbf{k} + \mathbf{q}/2,\uparrow} \Delta_{\mathbf{k}} d^\dagger_{-\mathbf{k} + \mathbf{q}/2,\downarrow}$ in $H_{1\text{b}}$, which we refer to as 1-$\vec{q}$ pairing. This breaks $C_{3z}$ symmetry, but we can still use the Chern number $C$ and adiabatic connectivity to the $\vec{q}=0$ states above to distinguish the pairing states \cite{TheEnergeticsPaper}.
Focusing on the fully gapped regime (see SI for nodal pairing), we present in \figref{fig:finite_q} the tunneling conductance for $\vec{q} \neq 0$. 
In the weak-tunneling regime, \figref{fig:finite_q}a, we see that the hard gap is reduced when $\vec{q}\neq 0$, whilst the overall region of LDOS suppression increases. This is related to the fact that the $\vec{q}$-induced shifts of the particle and hole Fermi surfaces in \figref{fig:geometry_and_crossover}b lead to an increase of ``particle-hole nesting'' (and, thus, of the gap) in some part of $\mathbf{k}$-space at the expense of other regions.
Another manifestation of finite $\vec{q}$ in \figref{fig:finite_q}a is the splitting of the Van Hove peaks into two less pronounced peaks. Here we chose the direction of $\mathbf{q}$ along the mirror-symmetry axes of the spectrum, which is energetically favored in most parameter regimes \cite{TheEnergeticsPaper}.

\begin{figure}[tb]
    \centering
    \includegraphics[width=0.47\textwidth]{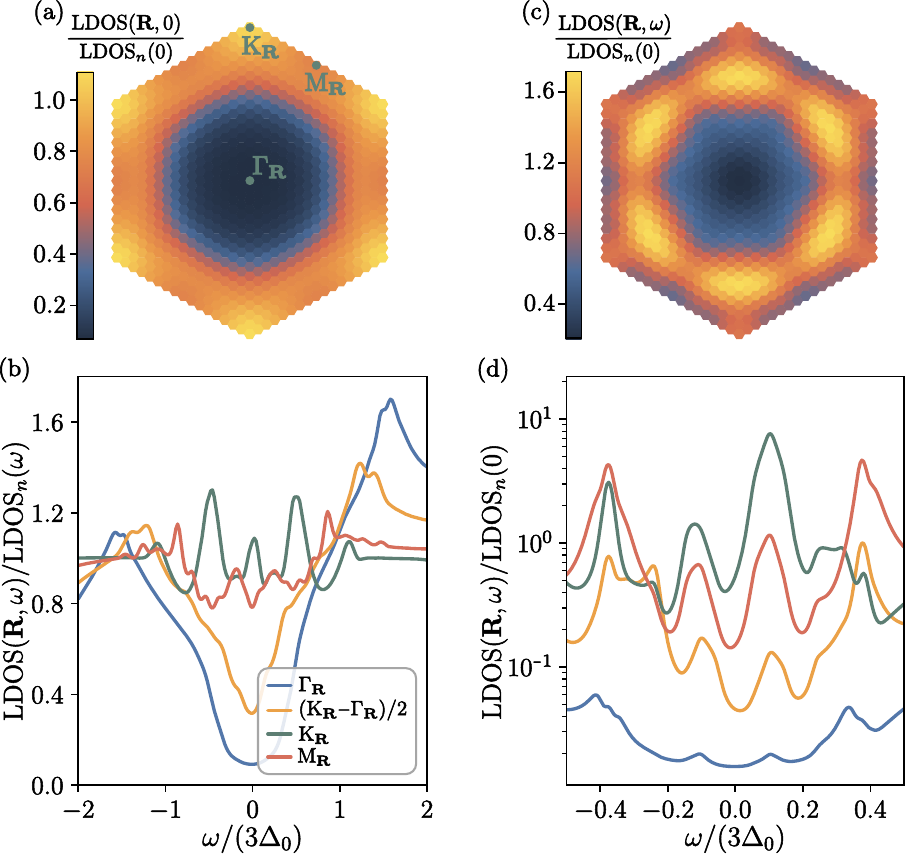}
    \caption{LDOS of the 3-$\vec{q}$ moiré state. a) Normalized LDOS at Fermi energy ($\omega = 0$) in the moiré unit cell in case of the nodal Bogoliubov excitations. b) Normalized LDOS at $\omega / (3\Delta_0) = 0.64$ in the moiré unit cell in case of the gapped Bogoliubov excitations. c) Energy dependence of LDOS in case of the nodal Bogoliubov at different points in the unit cell. d) Energy dependence of moiré Hamiltonian LDOS in case of the gapped Bogoliubov at different points in the unit cell. We here assume $\Delta_{\mathbf{k}} = \Delta_0$. In (a-b): $\mathbf{q}_1 / \braket{k_{\mathrm{F}}} =  (0.05,0)$, $\eta / (3\Delta_0) = 0.067 $; in (c-d): $\mathbf{q}_1 / \braket{k_{\mathrm{F}}} = (0.3,0), \eta / (3\Delta_0) = 0.017.$}
    \label{fig:three_q}
\end{figure}

In \figref{fig:finite_q}b, we compare the weak- and-strong tunneling conductance for different IRs of the order parameter and tunneling to the $A_1$ sublattice. As finite $\vec{q}$ breaks the $C_{3z}$ symmetry, it is expected to lead to a non-vanishing sub-gap Andreev current for all IRs. Even though the magnitude of $\vec{q}$ we chose leads to a significant reduction of the gap in the weak-tunneling limit, the qualitative behavior of the pairing is hardly affected for strong tunneling; we still see a significant difference between $A$ ($\Delta_{\mathbf{k}} = \Delta_0$) and $E$/$E^*$ ($\Delta_{\mathbf{k}} = \Delta_0 e^{\pm i\varphi_{\mathbf{k}}}$) such that the strong-tunneling regime can still be used to distinguish the three different pairing states with Chern numbers $C=0,\pm 1$.

\vspace{1em}
\textit{3-$\vec{q}$ superconductivity}---Despite the finite momentum of the superconductors discussed so far, lattice translation accompanied by an appropriate U(1) gauge transformation is still a symmetry such that any physical observable, like the tunneling conductance, will respect Bravais-lattice translational symmetry. However, it is also possible that a superposition of three $C_{3z}$-related momenta, $\mathbf{q}_j = C_{3z}^{j-1} \mathbf{q}$, $j=1,2,3$, is stabilized \cite{TheEnergeticsPaper}. This 3-$\vec{q}$ state is characterized by $H_{\Delta} = \sum_{\mathbf{k}} \sum_{j=1}^3 d_{\mathbf{k}+\mathbf{q}_{j}/2,\uparrow}^\dagger \Delta_{C_{3z}^{1-j}\mathbf{k}} d_{-\mathbf{k}+\mathbf{q}_{j}/2,\downarrow}^\dagger + \mathrm{H.c.}$. It respects $C_{3z}^M$ (modulo phase depending on $M$) but breaks translational symmetry explicitly and leads to a superlattice reconstruction of the continuum bands of RTG; the resulting model is analogous to the continuum model \cite{dos2007graphene,bistritzer2011moire} of moiré systems but with anomalous momentum-mixing terms.

Since this ``moiré superconductor'' can be distinguished from the states discussed earlier by its spatial variation, we only focus on the calculation of the LDOS here, which can be probed in the weak-tunneling regime. 
In \figref{fig:three_q}a, we present the LDOS (projected to the upper layer of RTG) in the emerging superlattice unit cell, at $\omega=0$ and for small $\vec{q}$ (much smaller than the Fermi wave vector). Intuitively, for small $\vec{q}$, one can think of the system at each position $\vec{R}$ locally as a superconductor with gap $\Delta_{\mathbf{k}}(\mathbf{R}) = \Delta_{\mathbf{k}} \sum_j  e^{i \mathbf{q}_j \mathbf{R}}$. This explains why we obtain low LDOS and reduced low-energy spectral weight, see also \figref{fig:three_q}b, near $\Gamma_{\mathbf{R}}$ associated with constructive interference of the three phases $e^{i \mathbf{q}_j \mathbf{R}}$ in $\Delta_{\mathbf{k}}(\mathbf{R})$; the destructive interference, on the other hand, leads to a lack of suppression of the LDOS at $\textrm{K}_{\mathbf{R}}$ and $\textrm{M}_{\mathbf{R}}$.

While there are some subtle differences in the LDOS between the slowly-varying local picture and the full model (see SI for details), additional features appear for larger $\vec{q}$, see \figref{fig:three_q}(c,d); this is also the energetically more favorable regime \cite{TheEnergeticsPaper} of the 3-$\vec{q}$ state. We can see in \figref{fig:three_q}(d) that the LDOS is more  suppressed at $\omega = 0$ for all positions (note the logarithmic scale). 
Above the gap, we see multiple complex peaks originating from the moiré reconstruction of the Bogoliubov bands and a non-trivial spatial dependence in the moiré unit cell, see \figref{fig:three_q}(c).

\vspace{1em}
\textit{Conclusion}---Our work demonstrates that the broken TRS in valley-imbalanced superconductors leads to a variety of features in tunneling spectroscopy that are different from pairing states emerging out of normal states with TRS. Given the remarkable progress in scanning tunneling studies of correlated graphene systems in recent years \cite{Oh_2021,TunnelingPerge,TIVCExp,WignerCrystals,OrderlyDisorder,2024arXiv241206476L,PhysRevB.91.035410,PhysRevB.107.L041404,2024arXiv241111163L}, we are confident that our formalism and results will provide crucial insights into the microscopic superconducting physics of RTG.

\begin{acknowledgments}
M.S.S. thanks Maine Christos and Pietro Bonetti for a collaboration on a related project on RTG \cite{TheEnergeticsPaper}.
All authors further acknowledge funding by the European Union (ERC-2021-STG, Project 101040651---SuperCorr). Views and opinions expressed are however those of the authors only and do not necessarily reflect those of the European Union or the European Research Council Executive Agency. Neither the European Union nor the granting authority can be held responsible for them. M.S.S.~is also greatful for support by grant NSF PHY-2309135 for his stay at the Kavli Institute for Theoretical Physics (KITP) where a part of the research was done.
\end{acknowledgments}

\bibliography{draft_1.bib}

\onecolumngrid

\begin{appendix}

\section{General formalism and current splitting}\label{appendix:general formalism}

We incorporate and generalize the Hamiltonian approach to tunneling between two systems based on the out-of-equilibrium Kedlysh formalism~\cite{hamiltonian_approach_to_NS}. Below we present a general framework to calculate the tunneling current between two systems. The Hamiltonian of the system comprising on the mean-field level the nature of the two coupled systems and tunneling between them can be generally written as follows
\begin{align}
    \hat{H} = \psi^\dagger \left( H_0 + H_t \right) \psi, \label{eq:t_hamiltonian}
\end{align}
where $\psi^\dagger = (f^\dagger, c^\dagger)$, and $f^\dagger$, $c^\dagger$ denote the full set of creation operators of the two systems; matrices $H_0, H_t$ describe uncoupled systems and tunneling between them, respectively, and are given as
\begin{align}
    H_0 =
    \begin{pmatrix}
        H_{f} & 0\\
        0 & H_{c}
    \end{pmatrix},\quad
    H_t =
    \begin{pmatrix}
        0 & T_t^\dagger\\
        T_t & 0
    \end{pmatrix}.
\end{align}
We assume that both systems are kept at the same temperature and the difference between their chemical potentials is given by a bias voltage, $\mu_f - \mu_c = eV$. Then the steady-state tunneling current reads
\begin{align}
    j(\tau) = -\sum_i e_{f,i} \braket{\dot{n}_{f,i}(\tau)} = \frac{i}{\hbar} \sum_{i,j} \left[ e_{f,i} (T^\dagger_t)_{ij} \braket{f_i^\dagger(\tau) c_{j}(\tau)} - e_{f,i} T_{t,ij} \braket{c_{j}^\dagger(\tau) f_{i}(\tau)} \right],
\end{align}
where $e_{f,i} = \pm e$ is the charge corresponding to the $i$-th degree of freedom of the system-$f$, and the average of the operators are given by the Green function on the Keldysh contour, $G^{+-}_{ij}(\tau,\tau) = i \braket{\psi_j^\dagger(\tau) \psi_i(\tau)}$. The tunneling current can be conveniently rewritten in the following form
\begin{align}
    j = \frac{e}{2\pi \hbar} \int d\omega \mathrm{Tr} \left[ J G^{+-} (\omega) \right],\quad J =
    \begin{pmatrix}
        0 & Q_{f} T_t^\dagger  \\
        - T_t Q_{f} & 0
    \end{pmatrix}, \label{eq:general_current}
\end{align}
where matrix $Q_{ij}^f = \delta_{ij} e_{f,j}$ is the charge matrix of the $f$-system. The bare $+-$ Green's function $g^{+-}(\omega) = 2\pi i \, \mathrm{diag}\left[ \rho_{f}(\omega) n_{\mathrm{F}}^{f}(\omega), \rho_{c}(\omega) n_{\mathrm{F}}^{c}(\omega) \right]$, where $\rho_{c,f}(\omega)$ are spectral functions of the systems, and $n_{\mathrm{F}}^{c,f}(\omega) = n_{\mathrm{F}}(\omega - Q_{c,f} \mu_{c,f})$, their thermal Fermi-Dirac distribution functions. The $G^{+-}$ is connected to $g^{+-}$ in the following way
\begin{align}
    G^{+-} = (1 + G^r H_t) g^{+-} (1 + H_t G^a)
\end{align}
where advanced and retarded Green functions $G^{a,r}$ satisfy a Dyson equation,
\begin{align}
    G^{a,r} = (1 + G^{a,r} H_t) g^{a,r} = g^{a,r} (1 + H_t G^{a,r}) \label{eq:dyson_eq}
\end{align}
with $g^{r,a}(\omega) = (\omega \pm i \eta - H_0 )^{-1}$ and $\eta$ being a finite lifetime broadening. 

This approach implicitly assumes that changes in the chemical potentials (and temperature) due to the tunneling are negligible, so that characteristic time scales of charging are much smaller than the time scales associated with $H_t$.

This formalism can be applied to a wide range of tunneling problems, herein we focus on the system modeling STS experiment, comprising a one-dimensional lead (its degrees of freedom are denoted by the $f$ operators) and superconductor (SC) of interest ($c$ operators). A general mean-field Hamiltonian of the system with opposite-spin pairing superconductivity can be written as follows
\begin{align}
    \hat{H}_{c} = \sum_{\mathbf{k}} (c^\dagger_{\mathbf{k},\uparrow} h_{\mathbf{k}} c_{\mathbf{k},\uparrow} - c_{-\mathbf{k},\downarrow} \cdot h_{-\mathbf{k}}^{\mathrm{T}} \cdot c^\dagger_{-\mathbf{k},\downarrow}) + \sum_{\mathbf{k},\mathbf{k}'} ( c^\dagger_{\mathbf{k},\uparrow} \cdot \Delta_{\mathbf{k},-\mathbf{k}'} \cdot c^{\dagger}_{-\mathbf{k}',\downarrow} + \mathrm{H.c.}),
\end{align}
where $h_{\mathbf{k}}$ is a spin-independent Hamiltonian matrix describing the (multiband) structure of the system, and $\Delta_{\mathbf{k},-\mathbf{k}'}$ is a superconducting (SC) order parameter matrix. We further simplify our analysis by considering a local tunneling approximation with respect to the lead degrees of freedom, the tunneling Hamiltonian then reads
\begin{align}
    \hat{H}_t =  \sum_{\mathbf{k},\sigma} f^{\dagger}_{\sigma}(z=0) t^{\dagger}_{\mathbf{k},\sigma} c_{\mathbf{k},\sigma} + \mathrm{H.c.}
\end{align}
where $f^\dagger(z=0)$ creates an electron at the tip of the lead, $t_{\mathbf{k}}^\dagger$ is a row-vector with the same number of component as in $c^\dagger_{\mathbf{k},\sigma}$. We also consider tunneling that preserves spin-rotational symmetry, $t_{\mathbf{k},\uparrow} = t_{\mathbf{k},\downarrow} = t_{\mathbf{k}}$. In order to incorporate the approach described earlier, we move to Nambu space and introduce particle and hole operators, 
\begin{align}
    c^\dagger_{\mathbf{k},\mathrm{p}} = c^\dagger_{\mathbf{k},\uparrow},\ c^\dagger_{\mathbf{k},\mathrm{h}} = c_{-\mathbf{k},\downarrow},\ f^\dagger_{\mathrm{p}} = f^\dagger_{\uparrow}(z=0),\ f^\dagger_{\mathrm{h}} = f_{\downarrow}(z=0).
\end{align}
The structure of the Dyson equation~\eqref{eq:dyson_eq} and expression for the tunneling current~\eqref{eq:general_current} allow to work with the Green function projected onto the subspace $\mathrm{im} H_t$. In this subspace, we arrange operators in the following manner, $f^\dagger = (f^\dagger_{\mathrm{p}}, f^\dagger_{\mathrm{h}})$, $c^\dagger = (\ldots, c^\dagger_{\mathbf{k},\mathrm{p}},\ldots,c^\dagger_{\mathbf{k},\mathrm{h}},\ldots)$, where $\ldots$ denotes going through all indices/operators in $\mathbf{k}$-space. Therefore, $Q_f = \sigma_z$, $Q_c = \sigma_z \otimes \mathds{1}_{\dim \mathcal{H}_{c,\uparrow}}$, and finally the tunneling matrix reads
\begin{align}
    T_t^\dagger = 
    \begin{pmatrix} 
        \ldots t^\dagger_{\mathbf{k}} \ldots & 0\\
        0 & \ldots -t^{\mathrm{T}}_{-\mathbf{k}} \ldots 
    \end{pmatrix}.
\end{align}
Moreover, in this subspace, we set $\rho_f(\omega) = \mathrm{diag}(\rho_{f,\mathrm{pp}}(\omega - eV), \rho_{f,\mathrm{hh}}(\omega - eV))$, where $\rho_{f,\mathrm{pp/hh}}(\omega \mp eV)$ is a bare particle (hole) local density of states at the tip of the lead. It is defined in such a way that $\rho_{f,\mathrm{pp}}(\omega)$ corresponds to the LDOS at $\mu_f$. Finally, let us clarify the notation for the Green functions matrix elements,
\begin{align}
    G_{cc,\alpha \beta}(\omega,\mathbf{k},\mathbf{k}') &= -i \Braket{ c_{\mathbf{k},\alpha} (\omega) c_{\mathbf{k}',\beta}^\dagger(\omega) },\\
    G_{ff,\alpha \beta}(\omega) &= -i \Braket{ f_{\alpha} (\omega) f_{\beta}^\dagger(\omega) },\\
    G_{cf, \alpha\beta}(\omega,\mathbf{k}) &= -i \Braket{ c_{\mathbf{k},\alpha} (\omega) f_{\beta}^\dagger(\omega) },
\end{align}
where $\alpha  = \mathrm{p,h}; \beta = \mathrm{p,h}$.

In general, the tunneling current can be splited into four distinct terms with different physical interpretations, $I = I_1 + I_2 + I_3 + I_{\mathrm{A}}$,
\begin{subequations}
    \begin{align}
    &I_1 = \frac{4\pi e}{\hbar} \int d\omega \left| 1 + \sum_{\vec{k}_1} t^{\dagger}_{\vec{k}_1} G^r_{cf,\mathrm{pp}}(\omega, \vec{k}_1) \right|^2 \rho_{f,\mathrm{pp}}(\omega - eV) \sum_{\vec{k},\vec{k}'} t^{\dagger}_{\vec{k}} \rho_{f,\mathrm{pp}}(\omega) t_{\vec{k}'} \left[ \nf(\omega - eV) - \nf(\omega) \right],\\
    &\begin{aligned}
        I_2 = - \frac{8\pi e}{h} \int d\omega & \mathrm{Re} \left[ \sum_{\vec{k}_1} G_{fc,\mathrm{hp}}^a(\omega,\vec{k}_1) t^{*}_{-\vec{k}_1} \left( 1 + \sum_{\vec{k}_2} t^\dagger_{\vec{k}_2} G^r_{cf,\mathrm{pp}}(\omega, \vec{k}_2)  \right) \sum_{\vec{k},\vec{k}'}t^{\mathrm{\dagger}}_{\vec{k}} \rho_{c,\mathrm{ph}}(\omega,\vec{k},\vec{k}') t_{\vec{k}'} \right]\\ &\times \rho_{f,\mathrm{pp}}(\omega - eV) \left[ \nf(\omega - eV) - \nf(\omega) \right],
    \end{aligned}\\
    &I_3 = \frac{4\pi e}{\hbar} \left| \sum_{\vec{k}_1} t^\dagger_{\vec{k}_1} (G^r_{cf,\mathrm{ph}}(\omega, \vec{k}_1) \right|^2 \rho_{f,\mathrm{pp}}(\omega - eV) \sum_{\vec{k},\vec{k}'} t^{\mathrm{T}}_{-\vec{k}} \rho_{c,\mathrm{hh}}(\omega,\vec{k},\vec{k}') t^*_{-\vec{k}'} \left[ \nf(\omega - eV) - \nf(\omega) \right],\\
    &I_{\mathrm{A}} = \frac{4\pi e}{\hbar} \int d\omega \left| \sum_{\vec{k},\vec{k}'} t^{\dagger}_{-\vec{k}} G_{cc,\mathrm{ph}}^r(\omega,\vec{k},\vec{k}') t^*_{-\vec{k}'} \right|^2 \rho_{f,\mathrm{pp}}(\omega - eV) \rho_{f,\mathrm{hh}}(\omega + eV) \left[ \nf(\omega - eV) - \nf(\omega + eV) \right]. \label{eq:Andreev_current_SI}
\end{align}
\end{subequations}

We are primarily interested in two terms that allow one to extract most of the information about the spectral properties of the Bogoliubov excitations and symmetry of superconductivity in the weak and strong tunneling regimes. The weak tunneling, when tunneling amplitudes are much smaller than characteristic energies of the systems (e.g. bandwidth, hopping constants), is mostly determined by the local density of states (LDOS), $I_{\text{weak}} \sim \int d\omega \rho_{f,\text{pp}}(\omega - eV) \sum_{\mathbf{k},\vec{k}'} t^\dagger_{\mathbf{k}} \rho_{c,\text{pp}}(\omega, \mathbf{k}, \vec{k}') t_{\mathbf{k}'} \left[ \nf(\omega - eV) - \nf(\omega) \right] + o(\| t \|^2)$, it describes the transfer of an electron from lead to a superconductor. In the strong tunneling, when the tunneling amplitudes are comparable, the Andreev current $I_{\mathrm{A}}$ can become non-negligible. Andreev current is generated by the process in which an electron tunnels from the lead to the SC, then by the Andreev scattering is converted to the hole tunneling back to the lead. This is the only that does not vanish when $|eV|$ lies with the gap of the Bogoliubov excitations, and as we show later it can be useful for probing the symmetry of the SC order parameter.

With some slight modifications, this approach can also be applied to study the tunneling current in the spin-polarized system. The difference is that we should rewrite the problem in the extended Majorana basis,
\begin{align}
    \hat{H} = \frac{1}{2} \Psi^\dagger \mathcal{H} \Psi,\quad \Psi = (\psi, \psi^\dagger)^{\mathrm{T}}.
\end{align}
The dynamics of the systems is then described by the retarded (advanced) Green function,
\begin{align}
    G^{r,a}(\omega) = \frac{1}{\omega \pm i\eta - \mathcal{H}}.
\end{align}
$G^{+-}$ is then constructed in the same way as described earlier. Once again identifying particle and hole operators (which are not independent now), $c^\dagger_{\mathbf{k},\mathrm{p}} = c^\dagger_{\mathbf{k}},\ c^\dagger_{\mathbf{k},\mathrm{h}} = c_{-\mathbf{k}},\ f^\dagger_{\mathrm{p}} = f^\dagger(0),\ f^\dagger_{\mathrm{h}} = f(0)$, one get the same expression for the tunneling current up to an additional factor $1/2$. 

In our numerical calculations, to remove all possible effects associated with the band structure of the lead, we assume that its DOS in the tip is constant, $\rho_{f,\mathrm{pp}}(\omega) = \rho_{f,\mathrm{hh}}(\omega) = \mathrm{const}(\omega)$.

    \section{Hamitlonian and three $C_{3z}$ symmetries}

\begin{figure}[b]
    \includegraphics[width=0.5\textwidth]{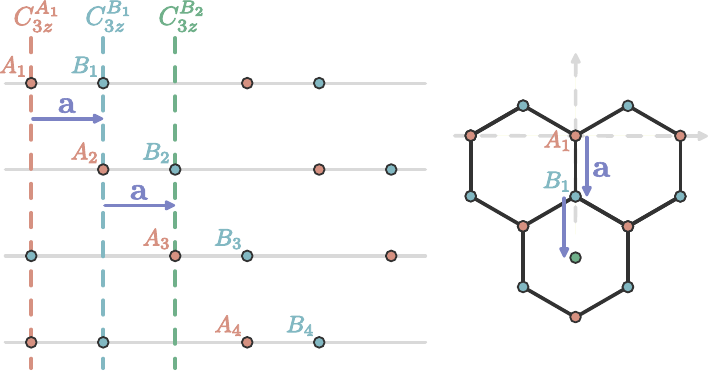}
    \caption{Geometry and $C_{3z}$-symmetry axes of the rhombohedral tetralayer graphene.}
    \label{fig:geometry_SI}
\end{figure}

ABCA-stacked rhombohedral tetralayer graphene consists of four stacked graphene monolayers each displaced relative to the previous one by the nearest neighbor lattice vector, its geometry is schematically depicted in Fig.~\ref{fig:geometry_SI} a). In our analysis, we use an 8-band $\mathbf{k}\cdot\mathbf{p}$ model
\begin{align}
    H=
    \begin{pmatrix} 
        u / 2 & v_0 \Pi^{\dagger} & v_4 \Pi^{\dagger} & v_3 \Pi & 0 & \gamma_2 / 2 & 0 & 0 \\
        v_0 \Pi & u / 2 & \gamma_1 & v_4 \Pi^{\dagger} & 0 & 0 & 0 & 0 \\
        v_4 \Pi & \gamma_1 & u / 6 & v_0 \Pi^{\dagger} & v_4 \Pi^{\dagger} & v_3 \Pi & 0 & \gamma_2 / 2 \\
        v_3 \Pi^{\dagger} & v_4 \Pi & v_0 \Pi & u / 6 & \gamma_1 & v_4 \Pi^{\dagger} & 0 & 0 \\
        0 & 0 & v_4 \Pi & \gamma_1 & -u / 6 & v_0 \Pi^{\dagger} & v_4 \Pi^{\dagger} & v_3 \Pi \\
        \gamma_2 / 2 & 0 & v_3 \Pi^{\dagger} & v_4 \Pi & v_0 \Pi & -u / 6 & \gamma_1 & v_4 \Pi^{\dagger} \\
        0 & 0 & 0 & 0 & v_4 \Pi & \gamma_1 & -u / 2 & v_0 \Pi^{\dagger} \\
        0 & 0 & \gamma_2 / 2 & 0 & v_3 \Pi^{\dagger} & v_4 \Pi & v_0 \Pi & -u / 2
    \end{pmatrix}
    \label{eq:rtg_hamiltonian_SI}
\end{align}
where $\Pi = \tau p_x +  i p_y$, $\tau = \pm $ corresponds to the valley, and $v_i = \sqrt{3} a \gamma_i / 2$ with graphene lattice constant $a = 0.246\, \text{nm}$. We used the parameters from~\cite{valleySCExp} $\gamma_0 = 3.1 \text{ eV}, \gamma_1 = 0.38 \text{ eV}, \gamma_2 = -0.0083 \text{ eV}, \gamma_3 = -0.29 \text{ eV}, \gamma_4 = -0.141 \text{ eV}$, and $u = 50\text{ meV}$ corresponding to the external potential.

There are three distinct $C_{3z}$ symmetries of the RTG which can be defined in real space as in-plain rotations by $2\pi/3$ around $A_1, B_1, B_2$, the rotation axes are shown in Fig.~\ref{fig:geometry_SI}. Let us closely study the action of the rotations on the periodic Bloch operators,
\begin{align}
    c^\dagger_{\mathbf{k},\alpha} = e^{-i \mathbf{k} \mathbf{a}_{\alpha} } \sum_{\mathbf{R}} e^{i \mathbf{k} ( \mathbf{R} + \mathbf{a}_{\alpha}) } c^\dagger_{\alpha}(\mathbf{R} + \mathbf{a}_{\alpha}),
\end{align}
where $\alpha = A_1, B_1, \ldots, A_4, B_4$, $\mathbf{a}_{A_n} = \mathbf{a} (n-1)//3, \mathbf{a}_{B_n} = \mathbf{a}n//3$, and $\mathbf{R}$ are Bravais lattice vectors. Choosing the origin of the coordinate system at $A_1$ the rotation of the operators reads
\begin{align}
    \begin{aligned}
        C_{3z}^{M}: c^{\dagger}_{\mathbf{k},\alpha}& \to  e^{-i \mathbf{ka}_{\alpha}} \sum_{\mathbf{R}} e^{i \mathbf{k} (\mathbf{R} + \mathbf{a}_{\alpha})} c_{\alpha}^\dagger(C_{3z}[\mathbf{R} + \mathbf{a}_{\alpha} - \mathbf{a}_M] + \mathbf{a}_M) \\
        &= e^{-i\mathbf{ka}_{\alpha}} \sum_{\mathbf{R}} \exp \left[ i \mathbf{k} \left\{ C_{3z}^{-1}(\mathbf{R} + \mathbf{a}_{\alpha} - \mathbf{a}_M) + \mathbf{a}_M \right\} \right] c_{\alpha}^\dagger(\mathbf{R} + \mathbf{a}_{\alpha})\\
        &= e^{-\mathbf{k}\mathbf{a}_{\alpha} } \sum_{\mathbf{R}} \exp \left[ i C_{3z} \mathbf{k} (\mathbf{R} + \mathbf{a}_{\alpha}) + i \mathbf{k} (\mathbf{a}_M - C_{3z}^{-1} \mathbf{a}_M) \right] c_{\alpha}^{\dagger}(\mathbf{R} + \mathbf{a}_{\alpha}),
    \end{aligned}
\end{align}
where $M = A_1, B_1, B_2$. Comparing this result with the expression for $c^\dagger_{C_{3z}\mathbf{k},\alpha}$, we obtain that the creation operators transform as follows 
\begin{align}
    C_{3z}^M: c^\dagger_{\mathbf{k},\alpha} \to \exp \left[ i (\mathbf{a}_M - \mathbf{a}_\alpha) (\mathbf{k} - C_{3z} \mathbf{k}) \right] c_{C_{3z}\mathbf{k},\alpha}^\dagger.
\end{align}
Let us now expand the transformation of the operators near $\mathbf{K}$ valley,
\begin{align}
    C_{3z}^M: c^\dagger_{\mathbf{K} + \mathbf{q},\alpha} \to e^{2\pi i /3 (\mathbf{a}_{\alpha} - \mathbf{a}_{M}) \cdot \mathbf{a} /|\mathbf{a}|^2 } \exp \left[ i (\mathbf{a}_M - \mathbf{a}_\alpha) (\mathbf{q} - C_{3z} \mathbf{q}) \right]  c_{\mathbf{K}+C_{3z}\mathbf{q},\alpha}^\dagger, 
\end{align}
We find it convenient to introduce a diagonal matrix corresponding to the transformation,
\begin{align}
    D^\dagger_{M,\alpha\alpha}(\mathbf{q}) =e^{2\pi i /3 (\mathbf{a}_{\alpha} - \mathbf{a}_{M}) \cdot \mathbf{a} /|\mathbf{a}|^2 } \exp \left[ i (\mathbf{a}_M - \mathbf{a}_\alpha) (\mathbf{q} - C_{3z} \mathbf{q}) \right];\quad  C_{3z}^M: c^{\dagger}_{\mathbf{K} + \mathbf{q}} \to c^\dagger_{\mathbf{K}+\mathbf{q}} D^\dagger_{M}(\mathbf{q}). \label{eq:operators_transformation_SI}
\end{align}
Later we consider $\mathbf{K}$-valley electrons and measure momentum from $\mathbf{K}$.

\section{Symmetry of the superconducting order parameter and Andreev current}

In this section, we consider the intra-band valley-polarized SC in RTG. Let $d_{\mathbf{k},\sigma}^\dagger = c^\dagger_{\mathbf{k},\sigma} u_{\mathbf{k}}$ be a fermionic operator corresponding to the band crossed by the chemical potential, where $u_{\mathbf{k},\sigma}$ is a correponding 8-component eigenvector of the matrix~\eqref{eq:rtg_hamiltonian_SI}. Let us choose a gauge in which the fermionic band operator does not acquire a phase under the rotation about $A_1$, $C_{3z}^{A_1}: d_{\mathbf{k},\sigma}^\dagger \to d_{C_{3z}\mathbf{k},\sigma}^\dagger$, then from~\eqref{eq:operators_transformation_SI}, we can identify the transformation properties of the $u_{\mathbf{k}}$,
\begin{align}
    C_{3z}^{A_1}: d_{\mathbf{k},\sigma}^\dagger &= c^{\dagger}_{\mathbf{k},\sigma} u_{\mathbf{k}} \to c^{\dagger}_{C_{3z}\mathbf{k},\sigma} D^{\dagger}_{A_1}(\mathbf{k}) u_{\mathbf{k}} = d_{C_{3z}\mathbf{k},\sigma}^\dagger = c^{\dagger}_{C_{3z}\mathbf{k},\sigma} u_{C_{3z}\mathbf{k}},\\
    u_{C_{3z}\mathbf{k}} &= D^{\dagger}_{A_1}(\mathbf{k}) u_{\mathbf{k}}. \label{eq:eigenvector_transformation_SI}
\end{align}
The SC order parameter is transformed as follows
\begin{align}
    &C_{3z}^{A_1}: \sum_{\mathbf{k}} d_{\mathbf{k},\uparrow}^\dagger \Delta_{\mathbf{k}} d_{-\mathbf{k},\downarrow}^\dagger \to \sum_{\mathbf{k}} d_{C_{3z} \mathbf{k},\uparrow}^\dagger \Delta_{\mathbf{k}} d_{-C_{3z} \mathbf{k},\downarrow}^\dagger = \sum_{\mathbf{k}} d_{\mathbf{k},\uparrow}^\dagger \Delta_{C_{3z}^{-1}\mathbf{k}} d_{-\mathbf{k},\downarrow}^\dagger,\\
    &C_{3z}^{A_1}: \Delta_{\mathbf{k}} \to \Delta_{C_{3z}^{-1}\mathbf{k}} = w \Delta_{\mathbf{k}},
\end{align}
where we associate a phase factor $w = 1, e^{2\pi i/3}, e^{-2\pi i/3}$ with the IRs $A, E, E^*$ of $C_{3z}^{A_1}$, respectively.

Using Eq.~(\ref{eq:operators_transformation_SI},~\ref{eq:eigenvector_transformation_SI}) we obtain the transformation of the band-operator under rotations around two other axes,
\begin{align}
   \begin{aligned}
        C_{3z}^{M}: d^\dagger_{\mathbf{k},\sigma} = c^\dagger_{\mathbf{k},\sigma} u_{\mathbf{k}} \to c^\dagger_{C_{3z}\mathbf{k},\sigma} D^{\dagger}_{M}(\mathbf{k}) u_{\mathbf{k}} &= c^\dagger_{C_{3z}\mathbf{k},\sigma} D^{\dagger}_{M}(\mathbf{k}) D_{A_1}(\mathbf{k}) u_{C_{3z}\mathbf{k}} = \gamma_M e^{i \mathbf{a}_{M}(\mathbf{k} - C_{3z}\mathbf{k})} c^\dagger_{C_{3z}\mathbf{k},\sigma} u_{C_{3z}\mathbf{k}}\\
        & = \gamma_M e^{i \mathbf{a}_{M}(\mathbf{k} - C_{3z}\mathbf{k})} d_{C_3z \mathbf{k},\sigma}^{\dagger},
   \end{aligned}
\end{align}
where we exploit the properties of the $D$ matrices: $ D^{\dagger}_{M}(\mathbf{k}) D_{A_1}(\mathbf{k}) = \gamma_M e^{i \mathbf{a}_{M}(\mathbf{k} - C_{3z}\mathbf{k})}$ with $\gamma_M = 1, e^{-2\pi i/3}, e^{2\pi i/3}$ for $M = A_1, B_1, B_2$. Therefore, the SC order parameter transforms as follows
\begin{align}
    &C_{3z}^{M}: \Delta_{k} \to \gamma_M^{2} \Delta_{C_{3z}^{-1}\mathbf{k}} = \gamma_M^2 w \Delta_{\mathbf{k}},
\end{align}
Thus, we obtain a unique property of the intra-valley SC order paramter: it transforms differently under different $C_{3z}$ rotations. This is in stark contrast to the intervalley SC, in which transformation of the order parameter is independent of the choice of the rotation axis due to the cancellation of the phase factors of the opposite momentum fermionic operators. Transformation of the order parameter for different $C_{3z}$-rotations is summarized in Table~\ref{table:irreps_SI}.

Now we focus on the selection rule for the Andreev current which can be formulated using the weight factor $|\sum_{\mathbf{k}} t^{\dagger}_{\mathbf{k}} G_{cc,\mathrm{ph}} (\omega, \mathbf{k},\mathbf{k}) t_{-\mathbf{k}}^*|$ entering the expression for the Andreev current. The tunneling Hamiltonian reads 
\begin{align}
    H_{\mathrm{tun}} =  \sum_{\mathbf{k},\alpha,\sigma} f^\dagger_{\sigma}(0) t_{\mathbf{k},\alpha}^{*} c_{\mathbf{k},\alpha,\sigma} + \text{H.c.}
\end{align}
where we again assumed preservation of the spin-rotational symmetry. Under rotations about $M$, the Hamiltonian transforms as follows
\begin{align}
    C_{3z}^{M}: H_{\mathrm{tun}} \to \sum_{\mathbf{k},\alpha,\sigma} f^\dagger_{\sigma}(0) t_{\mathbf{k}}^{\dagger} D_M(\mathbf{k}) c_{C_{3z}\mathbf{k},\sigma} + \text{H.c.} = \sum_{\mathbf{k},\alpha,\sigma} f^\dagger_{\sigma}(0) t_{C_{3z}^{-1}\mathbf{k}}^{\dagger} D_{M}(C_{3z}^{-1}\mathbf{k}) c_{\mathbf{k},\sigma} + \text{H.c.}
\end{align}
If $C_{3z}^{M}$ is preserved, we should require the following transformation of the tunneling coefficients $t^{\dagger}_{C_{3z} \mathbf{k}} D_{M}^\dagger(\mathbf{k}) = t^{\dagger}_{\mathbf{k}}$. We now study the weight-factor entering the expression for the Andreev current~\eqref{eq:Andreev_current_SI}, for simplicity, we replace the dressed Green function $G^r$ with the bare one,
\begin{align}
    \sum_{\mathbf{k}} t^{\dagger}_{\mathbf{k}} g_{cc,\mathrm{ph}} (\omega, \mathbf{k},\mathbf{k}) t_{-\mathbf{k}}^* =   \sum_{\mathbf{k}} \frac{1}{(\omega - \xi_{\mathbf{k}})(\omega + \xi_{-\mathbf{k}}) - |\Delta_{\mathbf{k}}|^2} \Delta_{\mathbf{k}} t_{\mathbf{k}}^{\dagger} u_{\mathbf{k}} t_{-\mathbf{k}}^{\dagger} u_{-\mathbf{k}} \label{eq:andreev_tunneling_weight_analysis_SI}
\end{align}
We first note that $[(\omega - \xi_{\mathbf{k}})(\omega + \xi_{-\mathbf{k}}) - |\Delta_{\mathbf{k}}|^2]^{-1}$ is symmetric under $\vec{k} \to C_{3z}\vec{k}$. Let us look closely at the function $V_{\mathbf{k}} = \Delta_{\mathbf{k}} t_{\mathbf{k}}^{\mathrm{T}} u_{\mathbf{k}} t_{-\mathbf{k}}^{\mathrm{T}} u_{-\mathbf{k}}$,
\begin{align}
    \begin{aligned}
        V_{C_{3z}\mathbf{k}} &= \Delta_{C_{3z} \mathbf{k}} t^\dagger_{C_{3z}\mathbf{k}} u_{C_{3z}\mathbf{k}} t^\dagger_{-C_{3z}\mathbf{k}} u_{-C_{3z}\mathbf{k}} \\
        &= w \Delta_{\mathbf{k}} t_{\mathbf{k}} D_{M}(\mathbf{k}) D^{\dagger}_{A_1}(\mathbf{k}) u_{\mathbf{k}} t_{\mathbf{k}} D_{M}(-\mathbf{k}) D^{\dagger}_{A_1}(-\mathbf{k}) u_{-\mathbf{k}}\\
        &= w^{-1}\gamma_M^{-2} V_{\mathbf{k}}.
    \end{aligned}
\end{align}
Therefore, the summation over $\vec{k}$ in Eq.~\eqref{eq:andreev_tunneling_weight_analysis_SI} does not vanish only if $\gamma_M^{-2} w^{-1} = 1$ which is a condition for the SC order parameter to transform trivially under $C_{3z}^{M}$. Thus, we conclude that the tunneling preserving one of the $C_{3z}$ symmetries results in the non-vanishing Andreev current only if the superconducting order parameter transforms under the trivial IR of the same symmetry. 


\begin{table}[t]
    \renewcommand{\arraystretch}{1.5}
    \begin{tabular}{|c|c|c|c|c|}
        \hline
        $\Delta_{\vec{k}}$ & $C_{3z}^{A_1}$ & $C_{3z}^{B_1}$ & $C_{3z}^{B_2}$ & $C$  \\ \hline
        $\Delta e^{i3m \varphi_{\vec{k}}}$  & $A$ & $E$ &  $E^*$ & $3m$ \\ \hline
        $\Delta e^{i(3m-1) \varphi_{\vec{k}}}$ & $E$ & $E^*$ & $A$ & $3m-1$ \\ \hline
        $\Delta e^{i(3m+1) \varphi_{\vec{k}}}$ & $E^*$ & $A$ & $E$ & $3m+1$ \\ \hline
    \end{tabular}
    \caption{Superconducting order parameter ($m \in \mathbb{Z}$) in the gauge used in \equref{wavefunctions} and the associated IRs with respect to the rotation around different points of the lattice. In the last column, we indicate the Chern number, as discussed in \appref{TopologyAndGaugeChoice}. The order parameters used in the main text correspond to $m=0$, i.e., the lowest harmonic in each channel.}
    \label{table:irreps_SI}
\end{table}

\section{Spin-polarized SC}

\subsection{Intervalley spin-polarized SC}

Let us consider an intervalley intraband spin-polarized SC. The Hamiltonian of this system can be explicitly written in the Majorana basis as follows
\begin{align}
    \hat{H} &= \frac{1}{2} \sum_{\mathbf{k}} \left( \xi_{\mathbf{k}} d^{\dagger}_{\mathbf{k},+} d_{\mathbf{k},+} + \xi_{\mathbf{k}} d^{\dagger}_{-\mathbf{k},-} d_{-\mathbf{k},-} - \xi_{\mathbf{k}} d_{\mathbf{k},+} d^{\dagger}_{\mathbf{k},+} - \xi_{\mathbf{k}} d_{-\mathbf{k},-} d^{\dagger}_{-\mathbf{k},-} \right) \\
    &+ \frac{1}{2} \sum_{\mathbf{k}} \left( \Delta_{\mathbf{k}} d^\dagger_{\mathbf{k},+} d^\dagger_{-\mathbf{k},-} - \Delta_{\mathbf{k}} d^\dagger_{-\mathbf{k},-} d^\dagger_{\mathbf{k},+} + \mathbf{H.c.} \right) \\
    & = \frac{1}{2} \sum_{\mathbf{k}} \Psi_{\mathbf{k}}^\dagger \mathcal{H}(\mathbf{k}) \Psi_{\mathbf{k}},
\end{align}
where we assume that TRS is preserved, $\xi_{\mathbf{k},+} = \xi_{-\mathbf{k},-} = \xi_{\mathbf{k}}$; $\Psi_{\mathbf{k}} = \left( d_{\mathbf{k},+}, d_{-\mathbf{k},-}, d_{\mathbf{k},+}^\dagger, d_{-\mathbf{k},-}^\dagger \right)$; and the Hamiltonian matrix reads
\begin{align}
    \begin{aligned}
        \mathcal{H}(\mathbf{k}) =
        \begin{pmatrix} 
            \xi_{\mathbf{k}} & & & \Delta_{\mathbf{k}} \\
            & \xi_{\mathbf{k}} & -\Delta_{\mathbf{k}} &\\
            & - \Delta_{\mathbf{k}}^{*} & -\xi_{\mathbf{k}} &\\
            \Delta_{\mathbf{k}}^{*} &&& - \xi_{\mathbf{k}}
        \end{pmatrix}.
    \end{aligned}
\end{align}

The tunneling Hamiltonian reads
\begin{align}
    \hat{H}_t = f^\dagger(0) \sum_{\mathbf{k}} \left[ t^\dagger_{\mathbf{k},+} c_{\mathbf{k},+} + t^\dagger_{\mathbf{k},-} c_{\mathbf{k},-} \right] + \mathrm{H.c.},
\end{align}
where $d^\dagger_{\mathbf{k},\pm} = c^\dagger_{\mathbf{k},\pm} u_{\mathbf{k}}$. Assuming that TRS is preserved, we require $t^\dagger_{\mathbf{k},+} = t_{-\mathbf{k},-}^{\mathrm{T}} = t^\dagger_{\mathbf{k}}$, and $u_{\mathbf{k},+} = u_{-\mathbf{k},-}^*$. Then the expression for the Andreev current is given by the same expression as before,
\begin{align}
    I_{\mathrm{A}} \propto \int d\omega \left| \sum_{\mathbf{k}} (t^\dagger_{\mathbf{k}}\ t^{\mathrm{T}}_{\mathbf{k}}) G^r_{cc,\mathrm{ph}}(\omega,\mathbf{k},\mathbf{k}) \begin{pmatrix} t_{\mathbf{k}}^* \\ t_{\mathbf{k}} \end{pmatrix}  \right|^2 \left[ \nf(\omega - eV) + \nf(\omega + eV) \right].
\end{align}
Let us again look at the bare weight factor,
\begin{align}
    \sum_{\mathbf{k}} (t^\dagger_{\mathbf{k}}\ t^{\mathrm{T}}_{\mathbf{k}}) g_{cc,\mathrm{ph}}(\omega,\mathbf{k},\mathbf{k}) \begin{pmatrix} t_{\mathbf{k}}^* \\ t_{\mathbf{k}} \end{pmatrix} = 
    \sum_{\mathbf{k}} \frac{1}{(\omega - \xi_{\mathbf{k}}) (\omega + \xi_{\mathbf{k}}) - |\Delta_{\mathbf{k}}|^2}
    (t^\dagger_{\mathbf{k}}\ t^{\mathrm{T}}_{\mathbf{k}})
    \begin{pmatrix} 
    0 & \Delta_{\mathbf{k}} u_{\mathbf{k}}  u_{\mathbf{k}}^{\dagger}\\
    -\Delta_{\mathbf{k}} u_{\mathbf{k}}^*  u_{\mathbf{k}}^{\mathrm{T}} & 0
    \end{pmatrix}
    \begin{pmatrix} t_{\mathbf{k}}^* \\ t_{\mathbf{k}} \end{pmatrix}.
\end{align}
Since
\begin{align}
    (t^\dagger_{\mathbf{k}}\ t^{\mathrm{T}}_{\mathbf{k}})
    \begin{pmatrix}
        0 & \Delta_{\mathbf{k}} u_{\mathbf{k}}  u_{\mathbf{k}}^{\dagger}\\
        -\Delta_{\mathbf{k}} u_{\mathbf{k}}^*  u_{\mathbf{k}}^{\mathrm{T}} & 0
    \end{pmatrix}
    \begin{pmatrix} t_{\mathbf{k}}^* \\ t_{\mathbf{k}} \end{pmatrix} = \Delta_{\mathbf{k}} \left[ t_{\mathbf{k}}^\dagger u_{\mathbf{k}} u_{\mathbf{k}}^\dagger t_{\mathbf{k}} - t_{\mathbf{k}}^{\mathrm{T}} u_{\mathbf{k}}^* u^{\mathrm{T}}_{\mathbf{k}} t^*_{\mathbf{k}} \right] = 0.
\end{align}
the Andreev current vanishes for an arbitrary order parameter. This is a direct consequence of the TRS.

\subsection{Intravalley spin-polarzed SC}

Let us consider an intraband spin-polarized SC, its Hamiltonian reads
\begin{align}
    H_{\mathrm{1b}} = \sum_{\mathbf{k}} \xi_{\mathbf{k}} d^\dagger_{\mathbf{k}} d_{\mathbf{k}} + \sum_{\mathbf{k}} \left[ \Delta_{\mathbf{k}} d_{\mathbf{k}}^\dagger d_{-\mathbf{k}}^\dagger + \mathrm{H.c.} \right],
\end{align}
where $\Delta_{\mathbf{k}} = -\Delta_{-\mathbf{k}}$, $d_{\mathbf{k}}^\dagger = c^{\dagger}_{\mathbf{k}} u_{\mathbf{k}}$, the spin index is omitted. We again assume that tunneling preserves spin, and therefore, the whole system can be treated as spin-polarized. Then the tunneling Hamiltonian is given as follows
\begin{align}
    H_{t} = f^\dagger(0) \sum_{\mathbf{k}} t^\dagger_{\mathbf{k}} c_{\mathbf{k}} + \mathrm{H.c.}
\end{align}
In this case, as noted in the~\appref{appendix:general formalism}, it is convenient to work in the extended Majorana basis. Then the the Andreev current is given by the same expression as before,
\begin{align}
    I_{\mathrm{A}} \propto \int d\omega \left| \sum_{\mathbf{k}} t^\dagger_{\mathbf{k}} G_{cc,\mathrm{ph}}^r(\omega,\mathbf{k},\mathbf{k}) t^*_{-\mathbf{k}} \right|^2 \left[ \nf(\omega - eV) + \nf(\omega + eV) \right].
\end{align}
Let us analyze the weight factor using bare Green function,
\begin{align}
    \sum_{\mathbf{k}} t^{\dagger}_{\mathbf{k}} g_{cc,\mathrm{ph}} (\omega, \mathbf{k},\mathbf{k}) t_{-\mathbf{k}}^* &= \sum_{\mathbf{k}} \frac{1}{(\omega - \xi_{\mathbf{k}})(\omega + \xi_{-\mathbf{k}}) - |\Delta_{\mathbf{k}}|^2} \Delta_{\mathbf{k}} t_{\mathbf{k}}^{\dagger} u_{\mathbf{k}} t_{-\mathbf{k}}^{\dagger} u_{-\mathbf{k}}\\
    &= \frac{1}{2} \sum_{\mathbf{k}} \Delta_{\mathbf{k}} t_{\mathbf{k}}^{\dagger} u_{\mathbf{k}} t_{-\mathbf{k}}^{\dagger} u_{-\mathbf{k}} \left[ \frac{1}{(\omega - \xi_{\mathbf{k}})(\omega + \xi_{-\mathbf{k}}) - |\Delta_{\mathbf{k}}|^2} - (\mathbf{k} \to -\mathbf{k}) \right].
    \label{}
\end{align}
In contrast to the spin-polarizaed inter-valley pairing discussed above, this expresion and hence the Andreev current does not vanish since the TRS is broken, $\xi_{\mathbf{k}} \neq \xi_{-\mathbf{k}}$. However, it vanishes at $\omega = 0$ for an arbitrary $\Delta_{\mathbf{k}}$. This suggests that in the spin-polarized case, the Andreev part of the tunneling conductance is generally smaller than in the spin-singlet SC.

\section{Topology and gauge choice}\label{TopologyAndGaugeChoice}
In this appendix, we will provide more details on the different contributions to the Berry curvature stated in the main text. To keep it general for now, let us consider a multiband Bloch Hamiltonian $h_{\vec{k}}$ (with arbitrary number of bands, internal degrees of freedom etc.) and assume superconductivity with center of mass momentum $\vec{Q}$ emerges in one of its bands, with energies $\epsilon_{\vec{k}}$ and Bloch states $\ket{u_{\vec{k}}}$. On the mean-field level, this can be described by the multi-band model
\begin{equation}
    H_{\text{MB}} = \sum_{\vec{k}} \Psi_{\vec{k}}^\dagger \hat{h}_{\text{BdG}}(\vec{k}) \Psi_{\vec{k}}^\pdagger, \quad \hat{h}_{\text{BdG}}(\vec{k}) = \begin{pmatrix}
        h_{\vec{k}+\vec{Q}/2} & \Delta_{\vec{k}} P \\ \Delta_{\vec{k}} P^\dagger & -h^*_{-\vec{k}+\vec{Q}/2},
    \end{pmatrix}, \quad \Psi_{\vec{k}} = \begin{pmatrix}
            c_{\vec{k}+\vec{Q}/2,\uparrow} \\ c^\dagger_{-\vec{k}+\vec{Q}/2,\downarrow}
        \end{pmatrix}, \label{BdGHamiltonian}
\end{equation}
where $c_{\vec{k},\sigma}$ are the electronic annihilation operators, $\Delta_{\vec{k}} \in \mathbb{C}$ the superconducting order parameter, and $P = \ket{u_{\vec{k}+\vec{Q}/2}}\bra{u^*_{-\vec{k}+\vec{Q}/2}}$ ensures that pairing only takes place in the aforementioned band. Our goal is to evaluate the Chern number of the occupied superconducting BdG band. From \equref{BdGHamiltonian}, we find that the associated wave function can be written as
\begin{equation}
    \ket{\Psi_{\vec{k}}} = \sum_{p=\pm} a_{\vec{k},p} \ket{\phi_{\vec{k},p}}\otimes \ket{p}_\tau, \quad \text{with} \quad \ket{\phi_{\vec{k},+}} = \ket{u_{\vec{k}+\vec{Q}/2}}, \,\, \ket{\phi_{\vec{k},-}} = \ket{u^*_{-\vec{k}+\vec{Q}/2}}, \label{BdGWavefunctions}
\end{equation}
where $\ket{p=+}_\tau = (1,0)^T$ and $\ket{p=-}_\tau = (0,1)^T$ are basis vectors in Nambu space. Furthermore, $\vec{a}_{\vec{k}} = (a_{\vec{k},+},a_{\vec{k},-})^T$ is the lower eigenvector of the effective one-band BdG Hamiltonian 
\begin{equation}
    h_{\text{BdG}}(\vec{k}) = \begin{pmatrix}
        \xi_{\vec{k} + \vec{Q}/2} & \Delta_{\vec{k}} \\ 
        \Delta_{\vec{k}}^* & -\xi_{-\vec{k} + \vec{Q}/2}
    \end{pmatrix} = E_0(\vec{k}) \tau_0 + \vec{g}_{\vec{k}} \cdot (\tau_x,\tau_y,\tau_z)^T, \label{EffBdGHam}
\end{equation}
which we also expanded in Pauli matrices $\tau_j$ for later reference. After straightforward algebra, one finds that the Berry curvature of the BdG wavefunctions in \equref{BdGWavefunctions} can be written as ($\partial_j = \partial / \partial{k_j}$)
\begin{subequations}
\begin{equation}
    \Omega(\vec{k}) \equiv  -2\,\Im \braket{\partial_{x}\Psi_{\vec{k}}| \partial_{y} \Psi_{\vec{k}}} = \Omega_{\text{SC}}(\vec{k}) + \Omega_{\text{b}}(\vec{k}) + \delta \Omega(\vec{k}), \label{SplittingOmegaInThreeParts}
\end{equation}
where 
\begin{equation}
    \Omega_{\text{SC}}(\vec{k}) = -2\,\Im [ (\partial_{x}\vec{a}^*_{\vec{k}}) \cdot \partial_{y} \vec{a}_{\vec{k}}] = \frac{1}{2 |\vec{g}_{\vec{k}}|^3} \vec{g}_{\vec{k}} \cdot (\partial_x \vec{g}_{\vec{k}} \times \partial_y \vec{g}_{\vec{k}})
\end{equation}
is the Berry curvature associated with the effective BdG Hamiltonian in \equref{EffBdGHam}, 
\begin{equation}
    \Omega_{\text{b}}(\vec{k}) = \sum_{p=\pm} |a_{\vec{k},p}|^2 p \, \Omega_u(p \vec{k} + \vec{Q}/2) , \quad \Omega_u(\vec{k})  = -2\,\Im \braket{\partial_{x}u_{\vec{k}}| \partial_{y} u_{\vec{k}}}
\end{equation}
is the (particle and hole) contribution from the Berry curvature $\Omega_u (\vec{k})$ of the bands hosting superconductivity, and 
\begin{equation}
    \delta \Omega(\vec{k}) = - 2 \, \Im \left[ \sum_{p=\pm} (\partial_{x}a^*_{\vec{k},p}) a_{\vec{k},p} \braket{\phi_{\vec{k},p}|\partial_y \phi_{\vec{k},p}} - (x \leftrightarrow y) \right]
\end{equation}\label{BCExpression}\end{subequations}
is a mixing term. Before proceeding further with the application to RTG, a few comments are in order: as immediately follows from its definition, the total Berry curvature $\Omega(\vec{k})$ is invariant under $\vec{k}$-dependent U(1) gauge transformations of $\vec{a}_{\vec{k}}$ and of the Bloch states 
we express superconductivity in; formally, these correspond to
\begin{align}
    U_a(1): \,\,\,\,\quad \vec{a}_{\vec{k}} \quad &\longrightarrow \quad e^{i \alpha_{\vec{k}}} \, \vec{a}_{\vec{k}}, \\
    U_u(1): \quad \ket{u_{\vec{k}}} \quad &\longrightarrow \quad e^{i \alpha_{\vec{k}}} \, \ket{u_{\vec{k}}}, \\
    a_{\vec{k},p} \quad &\longrightarrow \quad e^{-i p \alpha_{p \vec{k} + \vec{Q}/2}} a_{\vec{k},p},
\end{align}
respectively. While all three terms in \equref{SplittingOmegaInThreeParts} are separately invariant under $U_a(1)$, only $\Omega_{\text{b}}(\vec{k})$ is invariant under $U_u(1)$. In this sense, $\delta \Omega(\vec{k})$ is a necessary additional term accompanying $\Omega_{\text{SC}}(\vec{k})$ to ensure gauge invariance. This can be made more explicit by writing
\begin{equation}
    \Omega_{\text{SC}}(\vec{k}) + \delta \Omega(\vec{k}) = -2\,\Im [ (D_{x}\vec{a}^*_{\vec{k}}) \cdot D_{y} \vec{a}_{\vec{k}}], \quad D_j = \tau_0 \partial_j + \begin{pmatrix}
        \braket{\phi_{\vec{k},+}|\partial_y \phi_{\vec{k},+}} & 0 \\ 
        0 & \braket{\phi_{\vec{k},-}|\partial_y \phi_{\vec{k},-}}
    \end{pmatrix},
\end{equation}
where $D_j$ can be thought of as the gauge co-variant derivative with respect to $U_u(1)$.

It is also instructive to consider limiting cases. First, if $h_{\vec{k}}$ is just a trivial one-band model, i.e., $h_{\vec{k}} = \epsilon_{\vec{k}}$, the Bloch states have no internal structure and we get $\ket{\partial_j u_{\vec{k}}} = 0$ such that $\Omega(\vec{k}) = \Omega_{\text{SC}}(\vec{k}) = \frac{1}{2 |\vec{g}_{\vec{k}}|^3} \vec{g}_{\vec{k}} \cdot (\partial_x \vec{g}_{\vec{k}} \times \partial_y \vec{g}_{\vec{k}})$, recovering the well-known standard expression. Another simple limit is $\xi_{\vec{k}} < 0$ with $|\xi_{\vec{k}}| \gg |\Delta_{\vec{k}}|$ for all momenta, corresponding to a fully filled band where superconductivity plays no role. It then holds $\vec{a}_{\vec{k}} \sim (1,0)^T$ and, thus, $\Omega(\vec{k}) = \Omega_u(\vec{k} + \vec{Q}/2)$, as it should be.

In the intravalley pairing regime of RTG discussed in the main text, we have $\vec{Q} = 2\vec{K} + \vec{q}$ (with $|\vec{q}| \ll |\vec{K}|$ or simply $\vec{q}=0$). The Chern number of superconducting band is then given by $C= \frac{1}{2\pi} \int_{\text{BZ}}  \diff^2 \vec{k} \, \Omega(\vec{k}) = \frac{1}{2\pi} \int_{\text{LE}}  \diff^2 \vec{k} \, \Omega(\vec{k}) + \frac{1}{2\pi} \int_{\text{BZ} \setminus \text{LE}}  \diff^2 \vec{k} \, \Omega(\vec{k})$, where we split the integral into the low-energy (LE) part, where the continuum model of the active valley used in this work applies (small $\xi_{\vec{K}+\vec{q}/2 \pm \vec{k}}$), and its complement. As we expect no superconductivity in the complement, we conclude from \equref{BCExpression} that $\Omega(\vec{k}) = - \Omega_u(-\vec{k}+\vec{K}+\vec{q}/2)$ for all $\vec{k} \in \text{BZ} \setminus \text{LE}$. Assuming that valley polarization only leads to an energetic deformation of the band but not to a topological phase transition in the normal state itself, TRS implies $\int_{\text{BZ}}  \diff^2 \vec{k} \, \Omega_u(-\vec{k}+\vec{K}+\vec{q}/2) = 0 $ and we, thus, get $\int_{\text{BZ} \setminus \text{LE}}  \diff^2 \vec{k} \, \Omega(\vec{k}) = \int_{\text{LE}}  \diff^2 \vec{k} \, \Omega_u(-\vec{k}+\vec{K}+\vec{q}/2)$. Taken together, we can write the superconducting Chern number entirely as a low-energy expression,
\begin{equation}
    C = \frac{1}{2\pi} \int_{\text{LE}}  \diff^2 \vec{k} \left[ \Omega_{\text{SC}}(\vec{k}) + \Omega_{\text{b}}(\vec{k}) + \delta \Omega(\vec{k}) + \Omega_u(-\vec{k}+\vec{K}+\vec{q}/2) \right], \label{ExpressionForChernNumberLE}
\end{equation}
which we can evaluate in the continuum model. To do this, let us for simplicity consider the minimal, isotropic two-band model of \equref{eq:rtg_hamiltonian_SI} where only the dominant degrees of freedom---the $A_1$ (upper component of Hamiltonian) and $B_4$ (lower component)---are kept \cite{Model_Koshino},
\begin{equation}
    h_{\vec{k}+\vec{K}} = \begin{pmatrix}
        u_0 - \mu & w_0 (k_x + i k_y)^4 \\ w_0 (k_x + i k_y)^4 & -u_0 - \mu
    \end{pmatrix}.
\end{equation}
A simple possible gauge choice for the wave functions in the upper band is given by
\begin{equation}
    \ket{u_{\vec{k}+\vec{K}}} = \mathcal{N}_{\vec{k}} \begin{pmatrix}
        \sqrt{u_0^2 + w_0 \vec{k}^8} + u_0 \\ w_0 (k_x + i k_y)^4
    \end{pmatrix}, \quad \mathcal{N}_{\vec{k}} > 0. \label{wavefunctions}
\end{equation}
In this gauge, $C_{3z}^{A_1}$ acts trivially, i.e., without an additional $\vec{k}$ dependent phase, on the electronic band creation operators $d^\dagger_{\vec{k}}$. Let us now assume $\vec{q}=0$ and take an order parameter $\Delta_{\vec{k}} = \Delta$, which thus transforms under $A$ of $C_{3z}^{A_1}$. Using the wavefunctions in \equref{wavefunctions}, one can finds
\begin{equation}
    \frac{1}{2\pi} \int_{\text{LE}}  \diff^2 \vec{k} \left[ \Omega_{\text{b}}(\vec{k}) + \delta \Omega(\vec{k}) + \Omega_u(-\vec{k}+\vec{K}+\vec{q}/2) \right] = 0. \label{Cancellation}
\end{equation}
In fact, this still holds if we generalize to $\Delta_{\vec{k}} = \Delta e^{i n \varphi_{\vec{k}}}$, $n \in\mathbb{Z}$, transforming under $E$ (for $n=3m-1$, $m\in\mathbb{Z}$), $E^*$ (for $n=3m+1$, $m\in\mathbb{Z}$), or $A$ ($n=3m$). This can be seen by noting that the Nambu components $\vec{a}^{(n)}_{\vec{k}}$ of the superconductor with $\Delta_{\vec{k}} = \Delta e^{i n \varphi_{\vec{k}}}$ can be related as $a^{(n)}_{\vec{k},+} = a^{(0)}_{\vec{k},+}$ and $a^{(n)}_{\vec{k},-} = e^{-i n\varphi_{\vec{k}}}a^{(0)}_{\vec{k},-}$; this additional $\vec{k}$-dependent phase factor is easily seen to not affect any of the terms in \equref{Cancellation}. Taken together, \equref{ExpressionForChernNumberLE} simplifies to 
\begin{equation}
    C = \frac{1}{2\pi} \int_{\text{LE}}  \diff^2 \vec{k}  \,\Omega_{\text{SC}}(\vec{k}) = \frac{1}{4\pi} \int_{\text{LE}}  \diff^2 \vec{k}  \,\frac{1}{|\vec{g}_{\vec{k}}|^3} \vec{g}_{\vec{k}} \cdot (\partial_x \vec{g}_{\vec{k}} \times \partial_y \vec{g}_{\vec{k}}) \label{FinalExpression}
\end{equation}
in this gauge, as stated in the main text. Although we derived this relation using constant $|\Delta_{\vec{k}}|$, it also holds exactly when $|\Delta_{\vec{k}}|$ depends on $\vec{k}$ since both \equref{FinalExpression} and \equref{ExpressionForChernNumberLE} can only change when the superconducting gap vanishes. The resulting Chern numbers are also indicated in \tableref{table:irreps_SI}.

\section{Finite momentum pairing results. Nodal case}

\begin{figure}[h]
    \centering
    \includegraphics[width=0.75\textwidth]{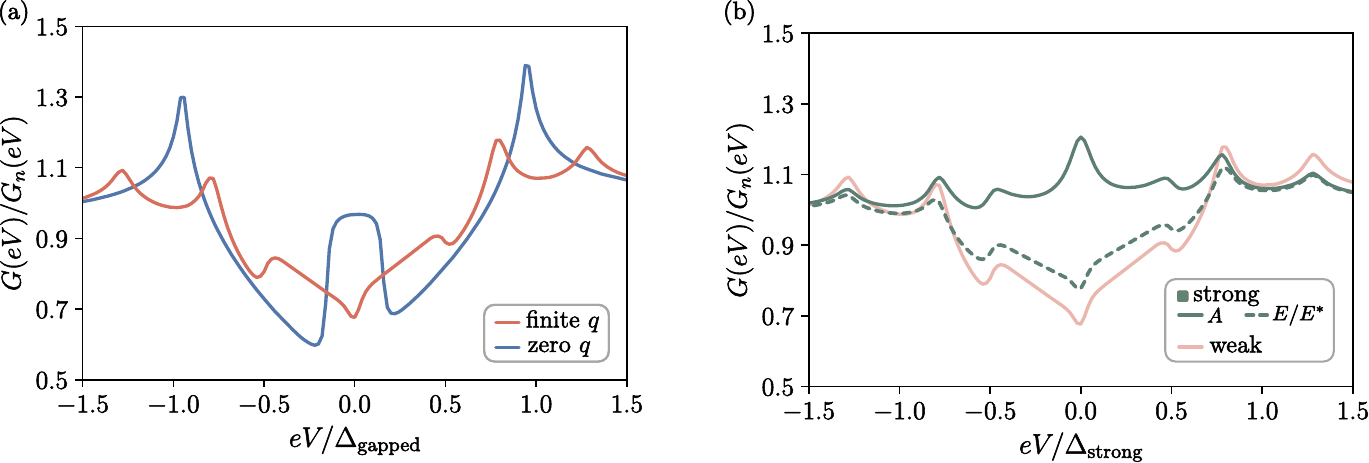}
    \caption{Tunneling conductance of the finite-momentum pairing in the nodal regime. a) Comparison of the weak tunneling conductance between zero-$q$ and finite-$q$ SC. b) Comparison between weak and strong tunneling conductance of the finite-$q$ SC for different IRs of the SC order parameter. Weak tunneling conductance is independent of the irrep of the order parameter.}
    \label{fig:finite_q_nodal_SI}
\end{figure}

Here we present complementary result for 1-$\vec{q}$ finite momentum pairing in the nodal regime discussed in the main text,
\begin{align}
    H_{\mathrm{SC}} = \sum_{\mathbf{k}} d^\dagger_{\mathbf{k} + \mathbf{q}/2,\uparrow} \Delta_{\mathbf{k}} d^\dagger_{-\mathbf{k} + \mathbf{q}/2,\downarrow} + \mathrm{H.c.}
\end{align}
In Fig.~\ref{fig:finite_q_nodal_SI}, we show weak and strong tunneling conductance for the different IRs of the order parameter in case of the nodal Bogolibov excitations. Here we use the same magnitude of SC order parameter $|\Delta_{\mathbf{k}}| = \Delta$ as in Fig. 2 of the main text.

\section{3-$\vec{q}$ moiré superconductivity. Slowly-varying approximation}

In this section, we present additional results for the 3-$\vec{q}$ moiré state discussed in the main text. The SC part of the Hamiltonian describing this state reads,
\begin{align}
    H_{\Delta} = \sum_{\mathbf{k}} \sum_{j=1}^3 d^\dagger_{\vec{k}+\vec{q}_j/2,\uparrow} \Delta_{C_{3z}^{1-j}\vec{k}} d^\dagger_{-\vec{k}+\vec{q}_j/2,\downarrow}.
\end{align}
Here we consider the $\vec{k}$-independent SC order parameter, $\Delta_{\vec{k}} = \Delta_0.$ Since this state can be distinguished from other states discussed in the paper by its spatial variation, we only focus on the calculation of the LDOS. In the limit of small $|\vec{q}|$, $|\vec{q}| / k_{\mathrm{F}} \ll 1$, one might expect that the intrinsic kinetic energy scales and the modulation of the SC order parameter can be decoupled, and therefore, at each point $\vec{R}$, we can locally approximate the SC Hamiltonian using the following one
\begin{align}
    H_{\Delta}(\vec{R}) = \Delta(\vec{R}) \sum_k \left( d^\dagger_{\vec{k},\uparrow} d^\dagger_{-\vec{k},\downarrow} + \mathrm{H.c.} \right)
\end{align}
where $\Delta(\vec{R}) = \Delta_0 \sum_j e^{i \vec{q}_j \vec{R}}$. We will call this approximation slowly-varying.

\begin{figure}
    \centering
    \includegraphics[width=1.0\linewidth]{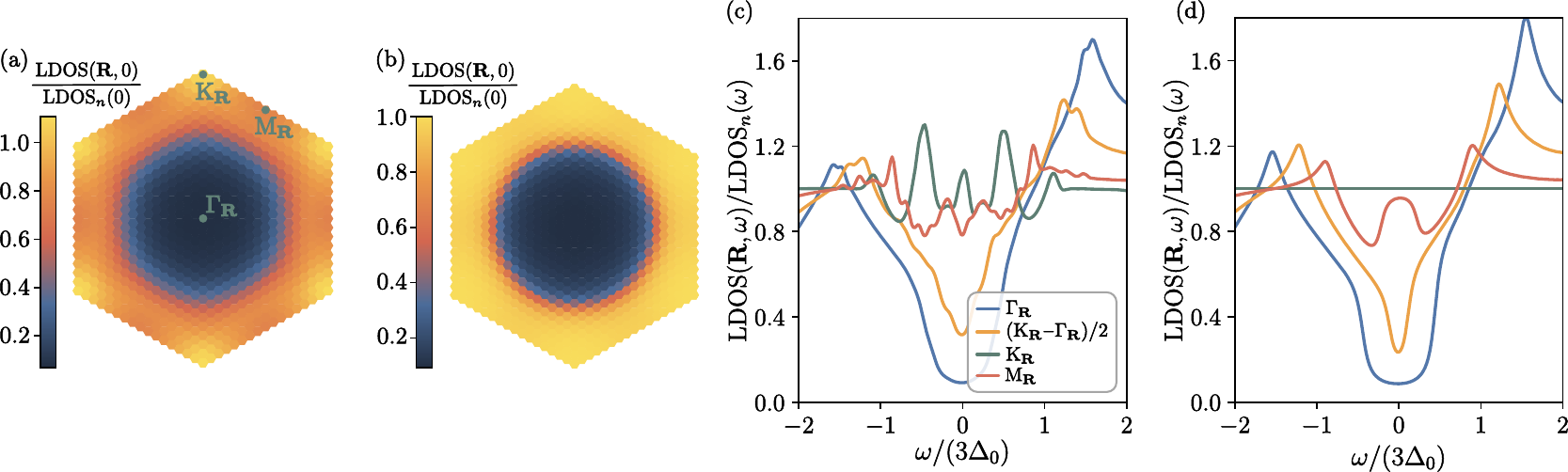}
    \caption{LDOS of the 3-$\vec{q}$ moiré SC state. a) Normalized LDOS at Fermi energy in the moirè unit cell found by numerical diagonalization of the moirè Hamiltonian. b) Normalized LDOS at Fermi energy in the moirè unit cell found in the slowly-varying approximation. c) Energy dependence of moirè Hamiltonian LDOS at different points in the unit cell. d) Energy dependence of LDOS in the slowly-varying approximation at different points in the unit cell.}
    \label{fig:three_q_SI}
\end{figure}

In Fig.~\ref{fig:three_q_SI}, we first present the LDOS projected to the upper layer of RTG at the Fermi surface in the emerging moiré unit cell found a) by the numerical diagonalization of the moiré Hamiltonian and b) in the slowly-varying approximation. In our calculations, we made $3|\Delta_0| = \max_{\vec{R}} |\Delta(\vec{R})| = |\Delta(0)|$ large enough, so that in the slowly-varying approximation at $\vec{R}=0$, the Bogoliubov excitations are gapped. In Fig.~\ref{fig:three_q_SI} (c-d), we support the comparison of two approaches by presenting the energy dependence of the LDOS at four points in the moirè unit cell. The intuitive slowly-varying limit gives a qualitatively good agreement with calculations based on the diagonalization of the Hamiltonian, however, the full LDOS has more structure, especially in the vicinity of the point $\mathrm{K}_{\mathbf{R}}$ where order parameter vanishes.

The limitations of this approximation are also illustrated in the next section using a 1D toy model.

\section{Moiré pairing in 1D toy model}

In this section, we discuss a pedagogical 1D toy-model with the SC order parameter that breaks translational invariance. We consider a one-band continuum model described by the Hamiltonian
\begin{align}
    H = \sum_{k,\sigma} \xi_k d^{\dagger}_{k,\sigma} d_{k,\sigma} + \frac{\Delta_0}{2} \sum_k \sum_{q = \pm q_0} \left( d^\dagger_{k + q/2,\uparrow} d^\dagger_{-k + q/2,\downarrow} + \mathrm{H.c.} \right), \label{eq:toy_model_hamiltonian_SI} 
\end{align}
without loss of generality we set $q_0 > 0$. To avoid divergencies associated with the normal DOS we take a linear dispersion $\xi_k = v_0 |k| - \mu$.

We are primarily interested in the LDOS. In the limit of small $q_0$, $v_0 q_0 / \mu \ll 1$, one might expect that the intrinsic kinetic energy scales and the modulation of the SC order parameter can be decoupled, and therefore, we can locally approximate the Hamiltonian with the following one
\begin{align}
    H_r = \sum_{k,\sigma} \xi_k d^{\dagger}_{k,\sigma} d_{k,\sigma} + \Delta(r) \sum_k \left( d^\dagger_{k,\uparrow} d^\dagger_{-k,\downarrow} + \mathrm{H.c.} \right)
\end{align}
where $\Delta(r) = \Delta_0 \cos(q_0 r)$.

\begin{figure}[h]
    \centering
    \includegraphics[width=0.55\textwidth]{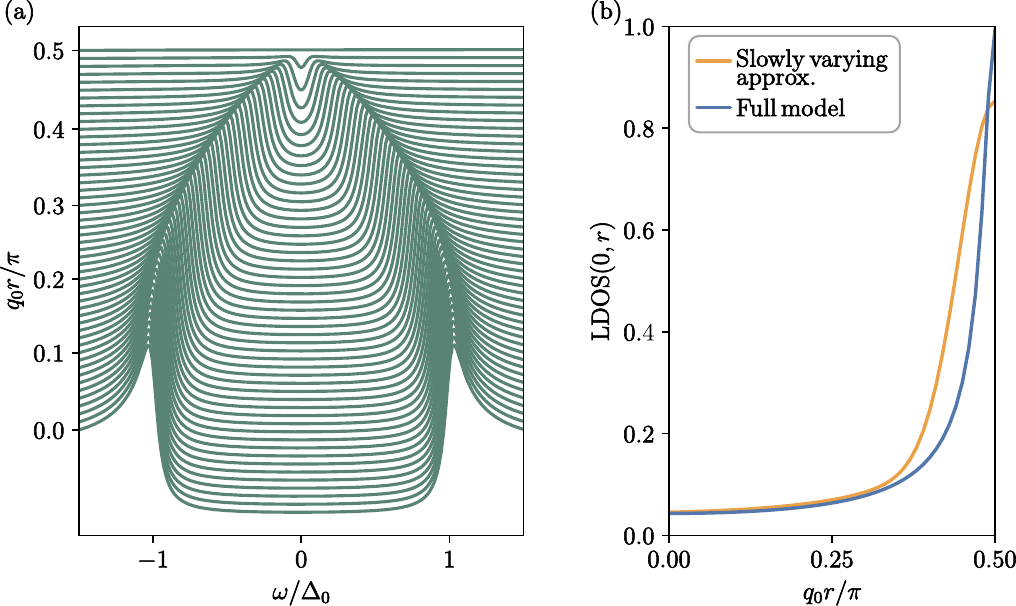}
    \caption{a) The LDOS in the slowly varying approximation. b) Comparison between LDOS found in the slowly varying approximation and by numerical diagonalization of the full model Hamiltonian. $v_0 q_0 /\mu = 0.01, \Delta / \mu = 0.4, \eta/\Delta_0 = 0.05$}
    \label{fig:toy_model_slowly_varying_SI}
\end{figure}

\begin{figure}[b]
    \centering
    \includegraphics[width=0.65\textwidth]{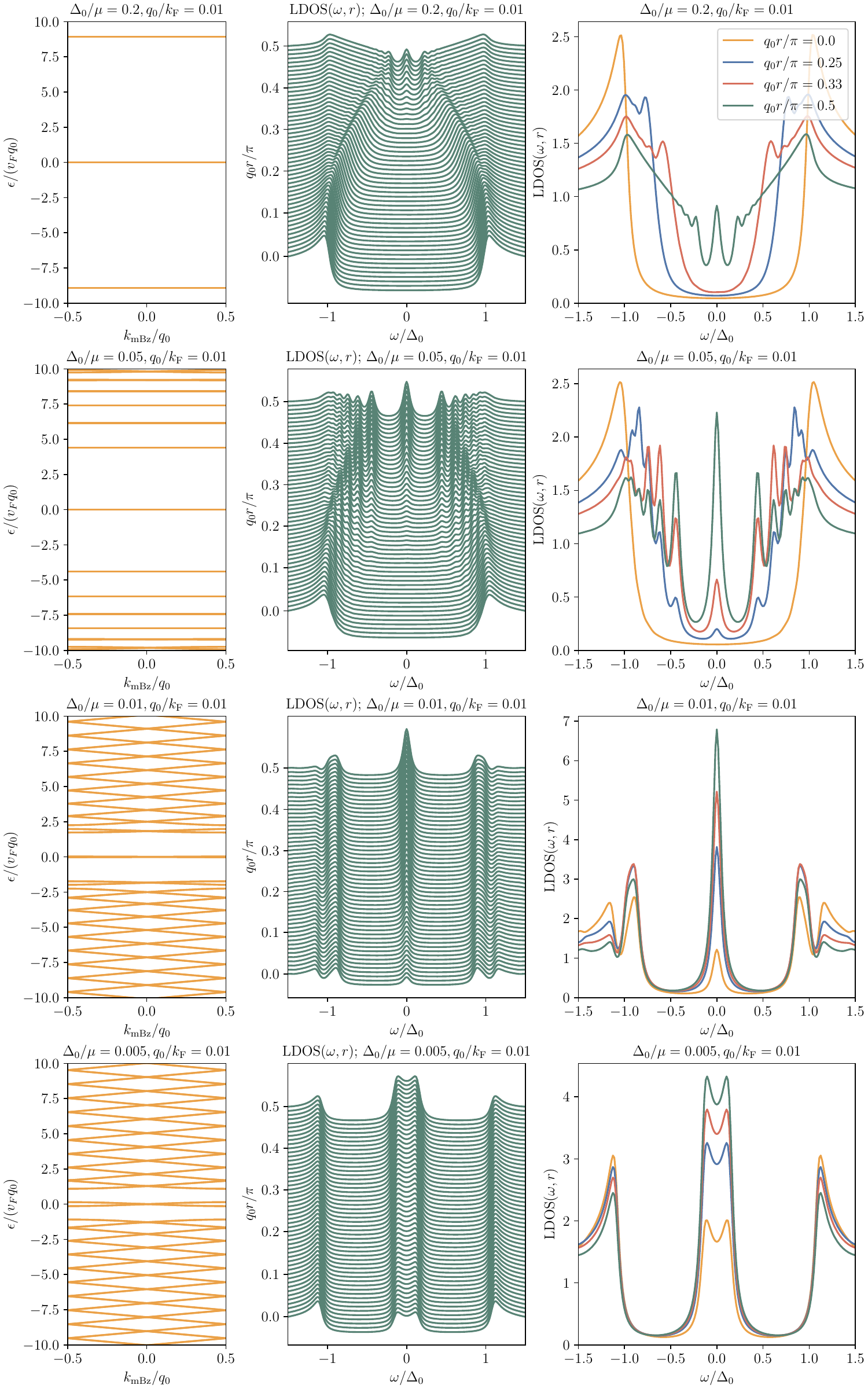}
    \caption{The overview of the moiré SC state in the 1D toy model in the limit $v_0 q_0 / \mu \ll 1$.}
    \label{fig:toy_model_overview_SI}
\end{figure}

In Fig.~\ref{fig:toy_model_slowly_varying_SI}a, we show the LDOS found in the slowly varying approximation for $v_0 q_0 / \mu = 0.01$. In Fig.~\ref{fig:toy_model_slowly_varying_SI}b, we compare the spatial modulation of the LDOS at zero frequency found in the slowly varying approximation and by the numerical diagonalization of the Hamiltonian~\eqref{eq:toy_model_hamiltonian_SI}. This comparison suggests that slowly varying approximation can be helpful in describing the LDOS of the system. However, further study of this model highlights the limitations of the approximation.

In Fig.~\ref{fig:toy_model_overview_SI}, we present a short overview of the SC moiré state for the same value of $v_0 q_0 / \mu = 0.01$. We first note that when the order parameter is much larger than $v_0 q_0$, the bands are almost flat. In the vicinity of $q_0 r = \pi/2$, where $\Delta(r)$ vanishes, the LDOS behaves unexpectedly: the peak at zero frequency surrounded by the two gaps appear. Furthermore, the gaps are followed by the linear increase of the DOS wich ends at $\omega/\Delta_0 \approx 1$. Decrease of $\Delta_0$ results in the more pronounced peak at zero frequency. Finally, in the limit of $\Delta_0 / v_0 q_0 \ll 1$ a clear two gap structure of the LDOS emerges for all $r$ in the unit cell.

\begin{figure}[t]
    \centering
    \includegraphics[width=1.0\textwidth]{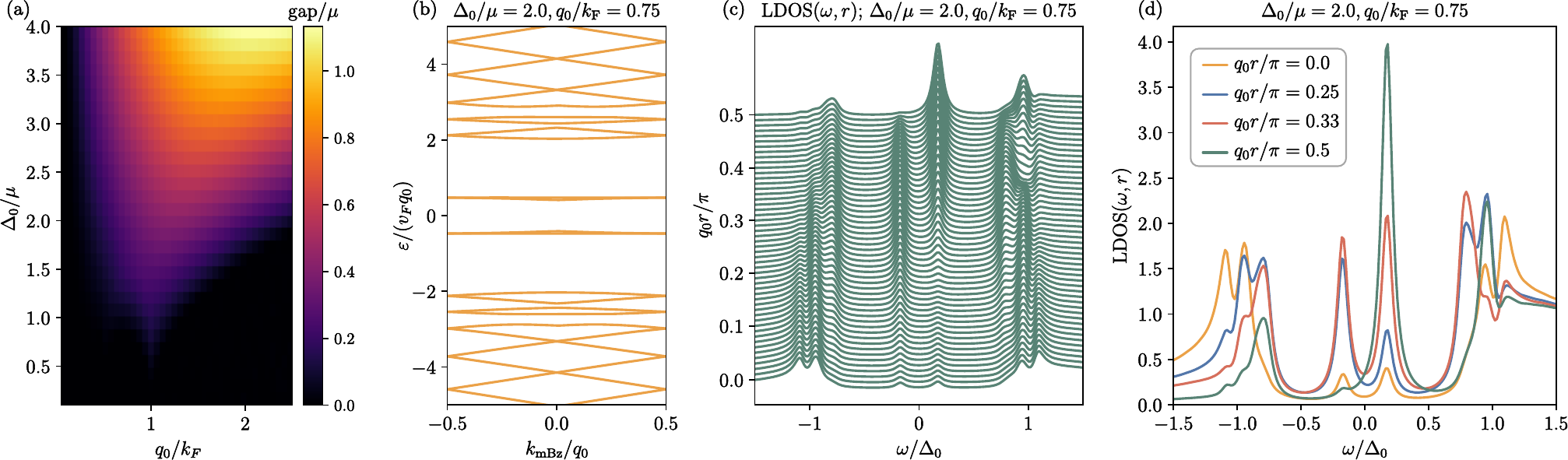}
    \caption{a) Bogoliubov excitation gap as a function of the magnitude of order parameter and pairing momentum. (b-d) The overview of the moiré SC state in the 1D toy model in fully-gapped limit.}
    \label{fig:toy_model_gapped_SI}
\end{figure}

We find that when $\Delta_0$ and $v_0 q_0$ are comparable to the chemical potential, Bogoliubov excitations are gapped. In Fig.~\ref{fig:toy_model_gapped_SI}a, we show the Bogoliubov excitation gap as a function of $\Delta_0$ and $v_0 q_0$. As an illustrative example, in Fig.~\ref{fig:toy_model_gapped_SI} (b-d) we also present the band structure and the LDOS for the parameters corresponding to the fully gapped regime.

Thus, we highlight the complexity of the moiré pairing and show that the slowly varying approximation gives a good agreement with the full model in the limit $v_0 q_0 /\mu \ll 1, \Delta_0 /\mu \ll 1, v_0 q_0 /\Delta_0 \ll 1$. However, this approximation cannot fully explain all the features of the LDOS.

\end{appendix}

\end{document}